%% file: main.tex
\documentclass[12pt,letterpaper,oneside]{article}

\usepackage[T1]{fontenc}

\usepackage{amsmath,amssymb,amsthm,mathtools}
\usepackage{newtxtext,newtxmath}
\usepackage{bm,mathrsfs}

\usepackage{booktabs,array,longtable,enumitem}
\usepackage{graphicx,placeins}
\usepackage{algorithm,algpseudocode}

\usepackage[authoryear,round]{natbib}

\input{layout.tex}

\usepackage{xcolor}
\definecolor{BGFLinkBlue}{HTML}{1765C1}
\usepackage[
  colorlinks=true,
  citecolor=BGFLinkBlue,
  linkcolor=BGFLinkBlue,
  urlcolor=BGFLinkBlue,
  filecolor=BGFLinkBlue
]{hyperref}
\usepackage[nameinlink,noabbrev]{cleveref}

\theoremstyle{compactplain}
\newtheorem{assumption}{Assumption}

\newtheorem{theorem}{Theorem}
\newtheorem{lemma}{Lemma}

\newtheorem{proposition}{Proposition}
\theoremstyle{compactremark}
\newtheorem{remark}{Remark}
\newtheorem{example}{Example}

\newcommand{\E}{\mathbb{E}}
\newcommand{\Prb}{\mathbb{P}}
\newcommand{\one}{\mathbf{1}}

\newcommand{\calW}{\mathcal{W}}

\newcommand{\calL}{\mathcal{L}}

\newcommand{\Kbm}{\mathbf{K}}

\newcommand{\Var}{\operatorname{Var}}
\DeclareMathOperator{\Cov}{Cov}
\newcommand{\MMD}{\operatorname{MMD}}
\newcommand{\argmin}{\operatorname*{arg\,min}}

\DeclareMathOperator{\clip}{clip}

\newcommand{\lead}[1]{
  \paragraph*{#1}
}

\hypersetup{
  pdftitle={Balance-Guided Forest for Causal Effect Estimation},
  pdfauthor={Feiyang Yi, Jelena Bradic}
}

\title{Balance-Guided Forest\\for Causal Effect Estimation}
\author{
  \begin{minipage}[t]{0.40\textwidth}
    \centering
    \textsc{Feiyang Yi}\\[0.45em]
    {\small Department of Statistics\\
    and Data Science\\
    Cornell University\\[0.25em]
    \textrm{fy238@cornell.edu}}
  \end{minipage}\hspace{0.02\textwidth}
  \begin{minipage}[t]{0.40\textwidth}
    \centering
    \textsc{Jelena Bradic}\\[0.45em]
    {\small Department of Statistics\\
    and Data Science\\
    Cornell University\\[0.25em]
    \textrm{jelena.bradic@cornell.edu}}
  \end{minipage}
}
\date{}

\begin{document}
\maketitle

\begin{abstract}
Covariate balancing is widely used for estimating causal effects in observational studies. A central challenge is determining which covariate features should be balanced. Forest-kernel balancing provides a data-driven way to learn informative balancing features. Existing forest-kernel balancing methods first learn a tree-based representation and then construct balancing weights. However, this procedure creates a mismatch, as limited tree capacity may be spent on features that are already well represented. We propose Balance-Guided Forest (BGF), which uses the current balancing weights to guide subsequent tree construction. BGF grows trees using a novel splitting score that combines residual imbalance with an outcome contrast, favoring splits that separate outcome-relevant regions where imbalance remains. This couples representation learning with covariate balancing rather than treating balancing as a downstream step. We establish an oracle-type bound that links the iterative construction to weighting error and causal bias. Simulations exhibit improvements in causal estimation relative to existing forest-kernel approaches. Real-data applications illustrate the practical behavior of BGF across observational studies.
\end{abstract}

\section{Introduction}
\label{sec:introduction}

In observational studies, differences in outcomes can reflect both treatment effects and differences in who receives treatment. Covariate balancing adjusts for measured differences by reweighting treatment groups to match selected functions of their covariates \citep{hainmueller2012,imai2014}. The choice of functions is consequential: the bias of a weighted outcome mean depends on imbalance in the conditional outcome mean \citep{kallus2020,brunssmith2022}. A representation can therefore be well balanced yet omit important confounding structure. A richer representation can capture this structure, but may require more variable weights \citep{zubizarreta2015,wang2020}. The challenge is to learn which functions need to be balanced.

Forest kernels learn such functions from tree partitions. Outcome-guided forests choose partitions for outcome prediction \citep{shen2026}, while joint treatment--outcome forests predict both responses \citep{de2025}. Both complete the forest before estimating the weights. \emph{A feature can remain predictive after it has been balanced.} A split with a smaller outcome contrast can then matter more for the remaining bias if its child proportions are still imbalanced. For a fixed forest size and depth, selecting already balanced features can leave other sources of bias unrepresented.

We propose \emph{Balance-Guided Forest} (BGF), which uses the current weights to decide which partition to add next. The criterion sums across arms the squared product of residual weighted imbalance and an outcome contrast. At the population level, each product is the bias contributed by the split's refinement of the outcome mean. After each tree, a capped quadratic program jointly recalibrates the weights against all accumulated leaves. The fitted outcome regressions remain fixed, while the discrepancies driving subsequent splits change. BGF therefore learns features according to their outcome relevance \emph{and} what previous weighting has already balanced.

\label{sec:related work}
This construction connects outcome-informed representation learning \citep{jin2025crossbalancing,assaad2021} with adaptive weighting. Boosted propensity estimation and adversarial weighting already adapt to discrepancies under the current weights \citep{zhao2019tailored,ozeryflato2020,kallus2020deepmatch}. BGF directs that search using the outcome contrast and refits the weights over all retained features. Its connection to boosting is through the residual guiding successive learners \citep{friedman2001boosting}. Here the residual is a weighted covariate discrepancy, and the update is joint calibration. Section~\ref{sec:bgf feedback} develops this connection for the actual refits.

For stump forests learned independently of the analysis sample, with additive outcome regressions and a uniform product covariate law, we derive a finite-sample risk bound separating sampling error, feature-balancing error, and residual confounding. The last admits a product bound in propensity and outcome approximation errors. Feedback controls expected scores averaged over late iterations. Under an additive step propensity and a signal condition on fitted outcome contrasts, this yields recovery of outcome-relevant assignment thresholds and ATE consistency.

Simulations hold the final weighting procedure fixed and isolate feedback through an ablation, with the largest gains for shallow forests. Two observational applications illustrate estimation and balance.

\section{Balancing Bias with Forests}
\label{sec:forest balancing}

\subsection{The role of the representation}
\label{sec:target}

Let $\{(X_i, A_i, Y_i)\}_{i=1}^n$ be i.i.d., with $X_i \in \mathcal{X}$, $A_i \in \{0, 1\}$, and $Y_i = Y_i(A_i)$. Write
\[
  \mu_a(x) = \E(Y \mid A = a, X = x)
  \quad\text{and}\quad
  \theta = \E\{Y(1) - Y(0)\}.
\] Let $Q$ be the marginal law of $X$, $P_a$ its law given $A = a$, and $Pf = \int f\,dP$.

\begin{assumption}[Identification]
\label{assump:identification}
\[
  \begin{aligned}
    \text{Consistency}\quad & Y = Y(A), \\
    \text{unconfoundedness}\quad & (Y(0), Y(1)) \perp A \mid X, \\
    \text{and overlap}\quad & 0 < \Pr(A = 1 \mid X) < 1
  \end{aligned}
\]
hold $Q$-almost surely.
\end{assumption}

Then $\theta = Q(\mu_1 - \mu_0)$ \citep{rosenbaum1983central}. For $w_a \ge 0$ with $P_a w_a = 1$, set $P_{a,w_a} f = P_a(w_a f)$ and $\theta(w) = P_{1,w_1} \mu_1 - P_{0,w_0} \mu_0$. The bias is
\begin{equation}
  \begin{aligned}
    \theta(w) - \theta = (P_{1,w_1} - Q) \mu_1 - (P_{0,w_0} - Q) \mu_0.
  \end{aligned}
  \label{eq:bias decomposition}
\end{equation}
The ideal weight $w_a^\star = \pi_a/e_a$, where $e_a(x) = \Pr(A = a \mid X = x)$ and $\pi_a = \Pr(A = a)$, balances every integrable function. Equation~\eqref{eq:bias decomposition} requires only balance of the two outcome means. This distinction motivates learning balancing features for the estimation problem.

For sample weights, let $\mathcal{I}_a = \{i:A_i = a\}$, $n_a = |\mathcal{I}_a|$, 
\[
  Q_n f = n^{-1} \sum_i f(X_i),
  \quad\text{and}\quad
  \widehat{P}_{a,w} f = \sum_{i \in \mathcal{I}_a} w_i f(X_i),
\]
with $w_i \ge 0$ and $\sum_{i \in \mathcal{I}_a} w_i = 1$. A representation $\Phi$ defines $\mathcal{F}_\Phi = \{c + \beta^\top \Phi:\|\beta\|_2 \le 1\}$. Write $r_a(w; \Phi) = (\widehat{P}_{a,w} - Q_n) \Phi$. Then
\begin{equation}
  \sup_{f \in \mathcal{F}_\Phi} |(\widehat{P}_{a,w} - Q_n)f| = \|r_a(w; \Phi)\|_2.
  \label{eq:function class imbalance}
\end{equation}
If $\mu_a = c + \beta^\top \Phi$ and analysis outcomes are excluded from weight fitting, the conditional bias relative to $Q_n \mu_a$ is $\beta^\top r_a$, and the conditional noise variance is $\sum_i w_i^2 \Var(Y_i \mid X_i, A_i = a)$. Thus $\|r_a\|_2^2 + \lambda\|w\|_2^2$ trades represented bias against a variance bound \citep{kallus2020,brunssmith2022}.

For any $g = c + \beta^\top \Phi$, an exact identity is
\begin{equation}
  \widehat{P}_{a,w} Y = Q_n g + \widehat{P}_{a,w}(Y - g) + \beta^\top r_a(w; \Phi).
  \label{eq:calibration regression}
\end{equation}
When $r_a = 0$, weighting agrees with a regression estimate plus a weighted residual correction, for every $g$ in the learned span. This is the model-assisted interpretation of calibration \citep{deville1992calibration,brunssmith2026regression}. Perturbing $g$ by $\varepsilon h$ changes the first two terms by $-\varepsilon(\widehat{P}_{a,w} - Q_n) h$. Balanced directions therefore give insensitivity to regression adjustment in those directions. For $\mu_a = g + u_a$, bias still includes $(\widehat{P}_{a,w} - Q_n) u_a$. A useful representation must control this omitted component as well as its measured imbalance \citep{clivio2024representation}.

\subsection{Forest features and adaptive balance}
\label{sec:forest kernels}

For probability laws $P, Q$ and a bounded positive semidefinite kernel $k$, let $\mathcal{H}_k$ be its reproducing kernel Hilbert space. The maximum mean discrepancy measures imbalance over the unit ball of $\mathcal{H}_k$: 
\begin{equation} 
  \MMD_k(P,Q) = \sup_{\|f\|_{\mathcal{H}_k}\le 1}|(P - Q)f|. 
\end{equation}
For a finite-feature kernel $k(x,x')=\Phi(x)^\top \Phi(x')$, this becomes $\MMD_k(P,Q)=\|(P - Q)\Phi\|_2$, recovering \eqref{eq:function class imbalance} for the empirical laws.

For a tree $T$ with terminal-leaf set $\calL(T)$, let $\phi_T(x) = (\one\{x \in \ell\})_{\ell \in \calL(T)}$ collect its terminal-leaf indicators. A forest defines
\begin{equation}
  \begin{aligned}
    \Phi_B(x) & = B^{-1/2}(\phi_{T_1}(x)^\top, \ldots, \phi_{T_B}(x)^\top)^\top, \\
    k_B(x, x') & = \langle \Phi_B(x), \Phi_B(x')\rangle.
  \end{aligned}
  \label{eq:forest kernel}
\end{equation}
Thus $k_B = B^{-1} \sum_b k_{T_b}$, with $k_T(x, x') = \langle \phi_T(x), \phi_T(x')\rangle$. The associated maximum mean discrepancy satisfies
\begin{equation}
  \MMD_{k_B}(\widehat{P}_{a,w}, Q_n) = \|r_a(w; \Phi_B)\|_2,
\end{equation}
and
\begin{equation}
  \begin{aligned}
    \|r_a(w; \Phi_B)\|_2^2 = \frac{1}{B} \sum_{b=1}^B \sum_{\ell \in \calL(T_b)} \{\widehat{P}_{a,w}(\ell) - Q_n(\ell)\}^2.
  \end{aligned}
  \label{eq:leaf balance}
\end{equation}
This is finite-feature kernel balancing \citep{wong2018,hazlett2020}. Trees determine both the balancing functions and their relative emphasis.

Outcome-guided forests \citep{shen2026} and joint treatment--outcome forests \citep{de2025} choose these features before weighting. Earlier adaptive methods include boosted propensity models that seek imbalanced trees \citep{zhao2019tailored} and adversarial methods that alternate discrepancy learning and weighting \citep{ozeryflato2020,kallus2020deepmatch}. BGF directs the search using the estimated bias contribution of each candidate split. Each weight refit then updates the residual imbalance used to score subsequent splits.

\section{Balance-Guided Forest}
\label{sec:bgf}

In this section, we describe the proposed Balance-Guided Forest (BGF), which alternates between growing trees with balance-guided splits and updating the weights through calibration. The fitted outcome regressions $\widehat{\mu}_0$ and $\widehat{\mu}_1$ remain fixed throughout forest construction, while the discrepancy used to select the next tree changes after each weight refit. We first derive the population splitting criterion for a single treatment arm, then give the ATE estimator based on a forest shared by both arms.

We use three independent samples, $\mathcal{D}_N, \mathcal{D}_S, \mathcal{D}_A$, for outcome regression fitting, forest construction, and final estimation, respectively. Cross-balancing similarly uses sample splitting to separate outcome-informed feature construction from weight estimation \citep{jin2025crossbalancing}. Here $\mathcal{D}_S$ also supports repeated weight feedback. Conditional on the constructed forest, the analysis stage has a fixed representation and its outcomes enter only the final weighted mean. During forest construction, $n$, $\mathcal{I}_a$, and $n_a$ refer to the structure sample $\mathcal{D}_S$.

\subsection{What a split contributes to bias}
\label{sec:population score}

Suppress the arm index and write $P_w f = P(wf)$, with $Pw = 1$. For $N = L \cup R$, let $q = Q(L \mid N) \in (0, 1)$ and $\mu_D = Q(\mu \mid D)$. Splitting $N$ replaces the parent mean $\mu_N$ with the child means $\mu_L$ and $\mu_R$. The resulting change in the outcome regression is
\begin{equation}
  \begin{aligned}
    h_{L;N} & = \mu_L \one_L + \mu_R \one_R - \mu_N \one_N\\
    & = (\mu_L - \mu_R)(\one_L - q \one_N).
  \end{aligned}
  \label{eq:centered refinement}
\end{equation}
Define the imbalance under the weight function $w$ as $\delta_w(L; N)=P_w(L)-qP_w(N)$. The contribution of this split to the weighting bias is $(P_w-Q)h_{L;N}=P_w h_{L;N}$, since $Qh_{L;N}=0$. We use the squared weighting bias contribution as the splitting score,
\begin{equation}
  \begin{aligned}
    G_w(L; N) & = \{(P_w - Q)h_{L;N}\}^2\\
    & = \delta_w(L; N)^2(\mu_L - \mu_R)^2.
  \end{aligned}
  \label{eq:population score}
\end{equation}

Normalize the splitting direction as
\[
  \psi_{L;N} = \frac{\one_L - q \one_N}{\{Q(N)q(1 - q)\}^{1/2}}.
\]
Then $Q \psi_{L;N} = 0$ and $Q \psi_{L;N}^2 = 1$. Let $\rho_w = dP_w/dQ - 1$. For arm $a$, $\rho_w = w e_a/\pi_a - 1$. When $\mu, \rho_w \in L_2(Q)$,
\begin{equation}
  \begin{aligned}
    h_{L;N} & = Q(\mu \psi_{L;N}) \psi_{L;N}, \\
    G_w(L; N) & = \{Q(\rho_w \psi_{L;N})Q(\mu \psi_{L;N})\}^2.
  \end{aligned}
  \label{eq:directional score}
\end{equation}
The score pairs weighting error and prognostic signal along the same direction. A large error orthogonal to the outcome has little bearing on bias. A strong outcome direction already balanced has little reason to be selected again.

For a tree rooted at $\mathcal{X}$, its normalized internal-node contrasts are orthonormal under $Q$. Define the projection $\Pi_T \mu = \sum_{\ell \in \mathcal{L}(T)} \mu_\ell \one_\ell$. The weighting bias for this projected outcome regression can be written as
\begin{equation}
  (P_w - Q) \Pi_T \mu = \sum_{N \in \mathcal{N}(T)} Q(\rho_w \psi_{L_N;N})Q(\mu \psi_{L_N;N}),
  \label{eq:tree bias expansion}
\end{equation}
where $\mathcal{N}(T)$ denotes the internal nodes, $L_N$ is the left child of $N$, and all leaves have positive $Q$-mass. These signed contributions can cancel when summed. BGF scores each split by the square of its contribution to this expansion.

The target-law prediction gain is $Qh_{L;N}^2 = Q(N)q(1 - q)(\mu_L - \mu_R)^2$. Hence
\begin{equation}
  G_w(L; N) = Qh_{L;N}^2\, \frac{\delta_w(L; N)^2}{Q(N)q(1 - q)}.
  \label{eq:cart factorization}
\end{equation}
Thus, BGF multiplies regression-tree prediction gain by standardized residual imbalance. Joint-response splitting instead combines prediction gains for treatment and outcome \citep{de2025}. BGF differs in that its splitting score accounts for the imbalance remaining \emph{after weighting}, allowing subsequent trees to focus on regions where imbalance remains.

\subsection{Balance-guided empirical splits}
\label{sec:bgf ate}

Fit $\widehat{\mu}_0,\widehat{\mu}_1$ on $\mathcal{D}_N$ and hold them fixed. Let $J$ index the current growth subsample of $\mathcal{D}_S$. On this subsample, for a region $D \subseteq\mathcal{X}$, define 
\[ 
  n(D)=\sum_{i\in J}\one\{X_i\in D\}, \qquad \bar{\mu}_{a,D}=\frac{1}{n(D)}\sum_{\substack{i\in J: X_i\in D}}\widehat{\mu}_a(X_i), \qquad \widehat{q}=\frac{n(L)}{n(N)}.
\] 
For current weights define 
\[
\omega_a(D)=\sum_{i\in J\cap\mathcal{I}_a}w_{a,i}\one\{X_i\in D\}, \qquad \widehat{\delta}_{a,w}(L;N)=\omega_a(L)-\widehat{q}\,\omega_a(N).
\]
These masses inherit the structure weights without renormalization. The empirical version of \eqref{eq:population score} gives the splitting criterion for ATE estimation as 
\begin{equation} 
  \widehat{G}_w^{\mathrm{ATE}}(L;N)=\sum_{a=0}^1\widehat{\delta}_{a,w}(L;N)^2(\bar{\mu}_{a,L}-\bar{\mu}_{a,R})^2. \label{eq:ate score} 
\end{equation}

For each $a$, the node mean $\bar{\mu}_{a,D}$ averages $\widehat{\mu}_a(X_i)$ over all growth observations in $D$, regardless of treatment assignment, since the population contrast is defined under $Q$. For $Q(D) > 0$, change of measure gives
\begin{equation}
  \begin{aligned}
    \E(Y \mid A = a, X \in D) - Q(\mu_a \mid D) = \frac{\Cov_Q(e_a, \mu_a \mid D)}{Q(e_a \mid D)}.
  \end{aligned}
  \label{eq:node selection covariance}
\end{equation}
Within-arm outcome means differ from target-law means by the conditional covariance term in \eqref{eq:node selection covariance}. The analogous covariance conditional on $\Phi_B(X)$ appears in the residual confounding term of Theorem~\ref{thm:bgf oracle}. Additive shifts of either fitted outcome regression leave the scores unchanged, and a common nonzero rescaling preserves their ranking.

\subsection{Calibrated weight feedback}
\label{sec:bgf feedback}

Initialize $w_{a,i}^{(0)} = 1/n_a$. After adding $T_t$, form $\Phi_t$ and its averaged kernel as in \eqref{eq:forest kernel}. With $C \ge 1$ and $\lambda_S > 0$, refit each arm by 
\begin{equation}
  \begin{aligned}
    w_a^{(t)} &= \argmin_{v \in \calW_a} \left\{\Bigl\|\sum_{i \in \mathcal{I}_a} v_i \Phi_t(X_i) - \frac{1}{n}\sum_{j=1}^n \Phi_t(X_j)\Bigr\|_2^2 + \lambda_S\|v\|_2^2\right\}, \\
    \calW_a &= \left\{v:\sum_{i \in \mathcal{I}_a} v_i=1,\quad 0\le v_i\le C/n_a\right\}.
  \end{aligned}
  \label{eq:ate refit}
\end{equation}
The refit recalibrates over all leaves of the trees built so far and updates the splitting scores in \eqref{eq:ate score} used to grow the next tree.

Define the current kernel discrepancy function
\[
  f_{a,t}(x) = \Phi_t(x)^\top r_a(w_a^{(t)}; \Phi_t).
\] For $C > 1$, the Karush--Kuhn--Tucker conditions give
\begin{equation}
  w_{a,i}^{(t)} = \clip_{[0, C/n_a]} \{c_{a,t} - f_{a,t}(X_i) / \lambda_S\},
  \label{eq:feedback witness}
\end{equation}
where $c_{a,t}$ enforces normalization. Each refit updates the weights using all leaves in the current forest, so adding a new tree can also change the coefficients of earlier leaf indicators in $f_{a,t}$. For $C = 1$, the only feasible weights are uniform. 

The effect is explicit for the first full-sample stump $H, H^c$. Let $\widehat{p}_a = n_a^{-1} \sum_{i \in \mathcal{I}_a} \one_H(X_i) \in (0, 1)$. If the cap is inactive,
\begin{equation}
  \widehat{P}_{a,w_a^{(1)}} H - Q_n H = \frac{\lambda_S}{2n_a \widehat{p}_a(1 - \widehat{p}_a) + \lambda_S} (\widehat{p}_a - Q_n H).
  \label{eq:stump feedback shrinkage}
\end{equation}
Thus the arm-specific score of that same split is multiplied by the square of the displayed shrinkage factor. An active cap can leave additional imbalance. Appendix~\ref{app:feedback identities} gives the exact constrained solution.

\Needspace{9\baselineskip}
\lead{Directions available to the next tree.}
On the full structure sample $\mathcal{D}_S$, fix an arm $a$ and let $Q_S$ denote the empirical law of $X$. Define $\rho_{a,t}=d\widehat P_{a,w_a^{(t)}}/dQ_S-1$ and $\mathcal{V}_t=\operatorname{span}\{1,\Phi_t\}$. Write $\langle f,g\rangle_{Q_S}=Q_S(fg)$ and $\|f\|_n^2 = Q_S f^2$, and let $\Pi_{\mathcal{V}_t}$ be the orthogonal projection onto $\mathcal{V}_t$ in $L_2(Q_S)$. For a normalized split contrast $\psi$, set $\psi_t^\perp = (I - \Pi_{\mathcal{V}_t})\psi$. Then
\begin{equation}
  \langle\rho_{a,t}, \psi\rangle_{Q_S} = \langle\rho_{a,t}, \Pi_{\mathcal{V}_t}\psi\rangle_{Q_S} + \langle(I-\Pi_{\mathcal{V}_t})\rho_{a,t},\psi_t^\perp\rangle_{Q_S}.
\end{equation}
Exact calibration makes the first term zero, so the discrepancy of a candidate split depends on its component outside the current leaf space $\mathcal{V}_t$. Ridge regularization and the weight cap can leave discrepancy within this space. The BGF splitting score uses the full discrepancy together with the estimated outcome contrast.

\begin{figure}[!tbp]
  \centering
  {\raggedright\small\textbf{(a)} Residual discrepancy\par}
  \vspace{2pt}
  \includegraphics[width=\linewidth]{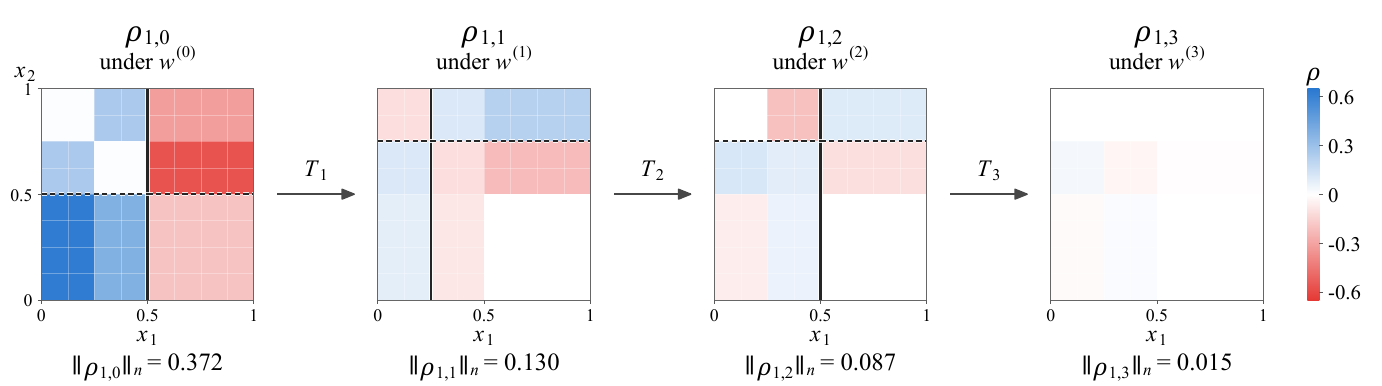}
  \par\vspace{7pt}
  {\raggedright\small\textbf{(b)} Selected trees\par}
  \vspace{2pt}
  \includegraphics[width=0.9\linewidth]{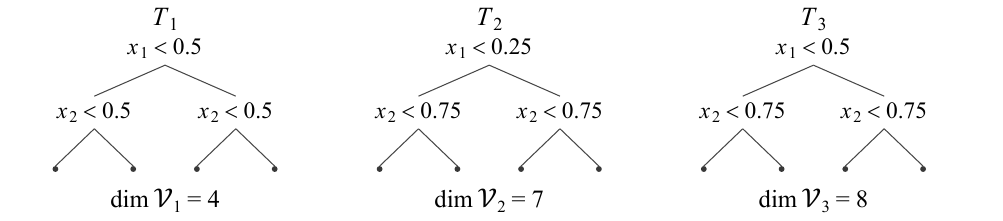}
  \caption{Weight feedback across successive BGF trees. (a) Treated-arm discrepancies under $w^{(t)}$, $t=0,1,2,3$. Arrows indicate the addition of a new tree followed by weight recalibration. Solid and dashed lines mark root and child splits of the next tree. (b) Selected trees and cumulative leaf-space dimensions.}
  \label{fig:weight feedback}
\end{figure}

Figure~\ref{fig:weight feedback} illustrates weight feedback across successive trees. The first three trees give $\dim\mathcal{V}_t=4,7,8$. The third tree adds $\one\{x_1<0.5\}\one\{x_2<0.75\}$ to $\mathcal{V}_2$. Previously used thresholds thus produce a new conditional direction. Weights remain fixed within each tree and are recalibrated over all leaves after growing a new tree. Appendix~\ref{app:feedback identities} specifies the construction details.

\lead{Boosting and Riesz regression.}
Squared-error boosting fits prediction residuals \citep{friedman2001boosting}. For $w_a, w_a^\star \in L_2(P_a)$, the Riesz loss $$\mathcal{R}_a(v) = P_a v^2 - 2Qv$$ has minimizer $w_a^\star$ \citep{chernozhukov2022riesznet} and derivative $D \mathcal{R}_a(w_a)[h] = 2(P_{a,w_a} - Q)h$. Hence
\begin{equation}
  G_{a,w_a}(L; N) = \tfrac{1}{4}\{D \mathcal{R}_a(w_a)[h_{a,L;N}]\}^2,
  \label{eq:riesz directional score}
\end{equation}
where $h_{a,L;N}$ is the outcome-scaled refinement in \eqref{eq:centered refinement}. RieszBoost updates the representer by gradient boosting \citep{lee2025rieszboost}. Tailored-loss boosting also targets imbalance \citep{zhao2019tailored}. For BGF, the derivative characterizes split selection. The updates jointly recalibrate weights over all leaves in the current forest by \eqref{eq:ate refit}, analogous to fully corrective boosting \citep{shen2013corrective}.

\lead{Iteration control.}
For a fixed arm, write $w_t = w_a^{(t)}$ and $$L_t(v) = \|r_a(v; \phi_{T_t})\|_2^2 + \lambda_S\|v\|_2^2.$$ The regularized follow-the-leader refits \citep{shalev2012online} satisfy, for adaptive trees and any $u \in \calW_a$,
\begin{equation}
  \begin{aligned}
    \sum_{t=1}^B L_t(w_t) + \lambda_S \sum_{t=2}^B (t - 1)\|w_t - w_{t-1}\|_2^2 \le \sum_{t=1}^B L_t(u).
  \end{aligned}
  \label{eq:feedback comparator budget}
\end{equation}
The comparator controls accumulated post-refit imbalance and weight movement, which bound pre-refit scores for full-sample stumps (Proposition~\ref{prop:bgf imbalance}).

\subsection{Other causal targets}
\label{sec:bgf extensions}

The same feedback construction applies when the target population or outcome functional changes. Let $Q^\dagger \ll P_a$ be a target covariate law supported by both arms, $b_a$ known bounded functions, and $z(Y)$ a transformed outcome. Consider
\[
  \begin{aligned}
    \Psi = \sum_{a=0}^1 Q^\dagger g_a, \qquad
    g_a(x) = b_a(x) \E\{z(Y) \mid A = a, X = x\}.
  \end{aligned}
\] 
Replace $Q$ by $Q^\dagger$ in the target feature moments, node proportions, and prediction averages, and use fitted estimates $\widehat g_a$ of $g_a$ in place of $\widehat\mu_a$. The split and refit rules are otherwise unchanged and only splits with $Q^\dagger(L) > 0$ and $Q^\dagger(R) > 0$ are considered. With weights calibrated to the empirical target law, estimate
\begin{equation}
  \widehat{\Psi} = \sum_{a=0}^1 \sum_{i \in \mathcal{I}_a} \widehat{w}_{a,i} b_a(X_i)z(Y_i).
  \label{eq:extended estimator}
\end{equation}
Its population weighting bias is $\sum_a (P_{a,w_a} - Q^\dagger)g_a$, so the same form of splitting score in \eqref{eq:population score} directs feature selection. 

For the effect on the treated, take $Q^\dagger = P_1$, $z(y) = y$, and $(b_1, b_0) = (1, -1)$. Treated weights remain uniform and control weights receive feedback. A prespecified subgroup instead uses $Q^\dagger = Q( \cdot \mid X \in H)$ \citep{li2018balancing}. For a fixed or independently learned policy $d$, use $Q^\dagger = Q$, $\widehat{g}_1 = d \widehat{\mu}_1$, and $\widehat {g}_0 = (1 - d)\widehat{\mu}_0$ \citep{kallus2018policy}. For distributional effects, use $z_u(y)=\one\{y\le u\}$ and set $\widehat g_a(x)=b_a(x)\widehat F_a(u\mid x)$, where $\widehat F_a(u\mid x)$ estimates $\Prb(Y \le u \mid A=a,X=x)$ \citep{firpo2007quantile}. Summing scores over a fixed grid gives common nonnegative weights, preserving monotonicity of each estimated distribution function.

\subsection{Estimation and implementation}
\label{sec:bgf algorithm}

At each node, offer $m_{\mathrm{try}}$ coordinates uniformly without replacement. Maximize \eqref{eq:ate score} over splits with at least $\max\{m, \lceil \alpha n(N)\rceil\}$ growth observations per child. Zero-score splits are allowed. Growth stops at depth $d_{\max}$ (root depth zero) or when no offered split is admissible.

Uniform weights are feasible for $C \ge 1$ and $\lambda_S > 0$ makes each refit unique. The cap implies effective sample size $1/\sum_i v_i^2 \ge n_a/C$. Repeated partitions are included as separate feature blocks and can change the relative penalties in \eqref{eq:leaf balance} even when the span is unchanged. 

After construction, refit on $\mathcal{D}_A$ using $\lambda_A$ to obtain $\widehat{w}_a$. The estimator is
\begin{equation}
  \widehat{\theta} = \sum_{i \in \mathcal{I}_1} \widehat{w}_{1,i} Y_i - \sum_{i \in \mathcal{I}_0} \widehat{w}_{0,i} Y_i.
  \label{eq:bgf ate}
\end{equation}

Algorithm~\ref{alg:bgf} summarizes the forest construction and final estimation procedure.

\begin{algorithm}
\caption{Balance-Guided Forest for ATE estimation}
\label{alg:bgf}
\begin{algorithmic}[1]
    \Statex \textbf{Input:} $\mathcal{D}_S, \mathcal{D}_A$; fitted outcome regressions $\widehat{\mu}_0, \widehat{\mu}_1$ from $\mathcal{D}_N$; $B$, $\lambda_S, \lambda_A > 0$, $C \ge 1$; growth parameters $(s, m_{\mathrm{try}}, m, \alpha, d_{\max})$.
    \State Initialize $w_0, w_1$ uniformly; set $k_0 = 0$.
    \For{$t = 1, \ldots, B$}
      \State Draw $s$ observations from $\mathcal{D}_S$ without replacement. Grow $T_t$ using \eqref{eq:ate score}, holding $w_0, w_1$ fixed.
      \State Set $k_t = \{(t - 1)k_{t-1} + k_{T_t}\}/t$. Refit both arms on all of $\mathcal{D}_S$ by \eqref{eq:ate refit}.
    \EndFor
    \State Freeze $k_B$. Refit weights on $\mathcal{D}_A$ with $\lambda_A$ and estimate $\theta$ by \eqref{eq:bgf ate}.
\end{algorithmic}
\end{algorithm}

\section{Theory}
\label{sec:theory}

In this section, we address three theoretical questions. First, what is the risk of balancing a fixed learned representation? Second, what does sequential weight feedback imply for the split scores that remain available? Third, when does this force recovery of confounding structure? Proofs appear in Appendix~\ref{app:theory proofs}.

\subsection{Risk of a learned representation}
\label{sec:theory oracle}

Let $\Phi_B$ be the representation in \eqref{eq:forest kernel}, with each tree a coordinate stump or a root-only block. For each stump, both leaf indicators are included, even if the stump duplicates an earlier one, and each block is scaled by $B^{-1/2}$. The forest is constructed independently of the i.i.d. analysis sample $\mathcal{D}_A$ of size $n$. Let $\mathcal{F}_B$ contain the construction information. Throughout this section, $\lambda_A$ and $\lambda_S$ are positive and deterministic. We set $\widehat{\theta}=0$ if either analysis arm is empty.

For any probability measure $P$, write $\|f\|_P^2 = Pf^2$. Unsubscripted vector norms are Euclidean, and $\| \cdot \|_2$ on $(0, 1)$ denotes Lebesgue norm.

\begin{assumption}[Overlap and additive outcomes]
\label{assump:theory regularity}
Assumption~\ref{assump:identification} holds, $p$ is fixed, and $Q$ is uniform on $[0, 1]^p$. For $a = 0, 1$, $\pi_a \ge \pi_*$ and $w_a^\star \le C_*$ $Q$-a.e., where $\pi_* > 0$ and $1 \le C_* < C$ are fixed. The outcome regressions satisfy
\[
  \mu_a(x) = c_a + \sum_{j=1}^p f_{aj}(x_j),
  \qquad
  R^2 := \sum_{a=0}^1 \sum_{j=1}^p \|f'_{aj}\|_2^2 < \infty,
\]
where each $f_{aj}$ is absolutely continuous and $\int_0^1 f_{aj} = 0$. Moreover, $|\mu_a| \le M$ and \mbox{$\Var(Y \mid X, A) \le \sigma_Y^2$}.
\end{assumption}

Fix the realized forest and let $X$ be a new covariate vector. Define 
\[ 
\begin{aligned} 
  \mu_{a,B}^Q(X)&=\E_Q\{\mu_a(X)\mid\Phi_B(X)\},\\ \mu_{a,B}^P(X)&=\E_{P_a}\{\mu_a(X)\mid\Phi_B(X)\},\\ e_{a,B}^Q(X)&=\E_Q\{e_a(X)\mid\Phi_B(X)\}, 
\end{aligned} 
\]
and write $\Delta_{a,B}(X)=\mu_{a,B}^P(X)-\mu_{a,B}^Q(X)$. We suppress the argument $X$ below.

\Needspace{7\baselineskip}
\begin{theorem}[Risk for a learned stump representation]
\label{thm:bgf oracle}
Under Assumption~\ref{assump:theory regularity}, let $\Phi_B$ be any measurable collection of $B$ coordinate stumps or root-only blocks constructed independently of $\mathcal{D}_A$. The estimator $\widehat{\theta}$ defined by Algorithm~\ref{alg:bgf} satisfies 
\begin{equation}
  \begin{aligned}
    \E\{(\widehat{\theta} - \theta)^2 \mid \mathcal{F}_B\}  \le K \left\{ \frac{M^2 + \sigma_Y^2 + BR^2}{n} + \sum_{a=0}^1 \big(P_a|\Delta_{a,B}|\big)^2 \right\}.
  \end{aligned}
  \label{eq:bgf oracle risk}
\end{equation}
Moreover,
\begin{equation}
  P_a|\Delta_{a,B}| \le \frac{1}{\pi_a} \|e_a - e_{a,B}^Q\|_Q \|\mu_a - \mu_{a,B}^Q\|_Q.
  \label{eq:bgf product remainder}
\end{equation}
The constant $K$ depends only on $(\pi_*, C_*, C, \lambda_A)$.
\end{theorem}

Theorem~\ref{thm:bgf oracle} applies to any stump forest constructed independently of the analysis sample. It separates estimation within the learned representation from residual confounding. Sampling contributes $(M^2 + \sigma_Y^2)/n$, and calibrating the represented outcome component contributes $BR^2/n$. In Algorithm~\ref{alg:bgf}, the independent analysis-sample refit controls these two terms. The product bound makes the remaining bias small when either the propensity score or the outcome regression is well represented. We next study how weight feedback guides the construction toward features that reduce residual confounding.

\begin{remark}[Bias decomposition for a fixed representation]
\label{rem:bgf bias}
For any weight function $w_a$ such that $w_a(X)$ is $\sigma\{\Phi_B(X)\}$-measurable, $P_a w_a=1$, and $0\le w_a\le C$,
\[
  \begin{aligned}
    (P_{a,w_a} - Q) \mu_a & = (P_{a,w_a} - Q) \mu_{a,B}^Q + P_{a,w_a} \Delta_{a,B}, \\
    |P_{a,w_a} \Delta_{a,B}| & \le C P_a|\Delta_{a,B}|.
  \end{aligned}
\]
Weight refitting addresses imbalance in the represented outcome component. The term $P_{a,w_a}\Delta_{a,B}$ captures residual confounding due to information omitted by the representation \citep{clivio2024representation}. Outcome-guided partition selection seeks to reduce this residual confounding.
\end{remark}

\subsection{Feedback and late-stage split scores}
\label{sec:theory feedback}

We now study the depth-one version of Algorithm~\ref{alg:bgf}. The nuisance, structure, and analysis samples have sizes $N, n, n$ and are independent. Each iteration adds one coordinate stump and refits on the full structure sample. Let $n_{S,a}$ be the sample size of arm $a$ in $\mathcal{D}_S$. Write $Q_S$ and $P_{S,a}$ for the empirical marginal and arm-specific covariate laws, and let $v_{a,t}(X_i) = n_{S,a} w_{a,i}^{(t)}.$ Then $P_{S,a} v_{a,t} = 1$ and $0 \le v_{a,t} \le C$.

For stump $b$, let $\phi_b$ be its two-leaf indicator vector and define
\[
  R_{a,b}(v) = P_{S,a}(v \phi_b) - Q_S \phi_b.
\]
Set
\begin{equation}
  \begin{aligned}
    \mathcal{Q}_t(v) = \frac{t \lambda_S}{n} \sum_{a=0}^1 P_{S,a} v_a^2 + \sum_{a=0}^1 \frac{n_{S,a}}{n} \sum_{b=1}^t \|R_{a,b}(v_a)\|^2.
  \end{aligned}
  \label{eq:feedback criterion}
\end{equation}
The weights $v_t = (v_{0,t}, v_{1,t})$ minimize $\mathcal{Q}_t$ subject to $P_{S,a}v_a=1$ and $0 \le v_a \le C$.

\begin{lemma}[Feedback energy inequality]
\label{lem:bgf descent}
Under Assumption~\ref{assump:theory regularity}, for every $1 \le t < B$,
\begin{equation}
  \begin{aligned}
    & \mathcal{Q}_{t+1}(v_{t+1}) - \mathcal{Q}_t(v_t) - \frac{2 \lambda_S}{n} \ge \frac{t \lambda_S}{n} \sum_{a=0}^1 \|v_{a,t+1} - v_{a,t}\|_{P_{S,a}}^2 + \sum_{a=0}^1 \frac{n_{S,a}}{n} \|R_{a,t+1}(v_{a,t+1})\|^2.
  \end{aligned}
  \label{eq:bgf descent}
\end{equation}
\end{lemma}

This lemma links each actual refit to two observable quantities: the movement of the weights and the imbalance remaining on the new leaves. Summing the inequality and bounding the terminal criterion by a feasible comparator controls their accumulation. In Algorithm~\ref{alg:bgf}, a large discrepancy selected before refitting must either be corrected through weight movement or remain in the new leaf residual. This is the mechanism used in the next result.

Assume further that $|\widehat{\mu}_a| \le G$ almost surely for some fixed $G < \infty$. Each iteration uses $s = n$, $d_{\max} = 1$, $\alpha = 0$, a fresh uniform subset of $m_{\mathrm{try}}$ coordinates, and at least $m$ observations per child. Let $S_{t+1}$ be the actual BGF score \eqref{eq:ate score} selected at iteration $t + 1$.

\begin{proposition}[Late-stage split-score control]
\label{prop:bgf imbalance}
Under Assumption~\ref{assump:theory regularity}, there is a constant $K$, depending only on $(p, \pi_*, C_*, C)$, such that for every $2 \le T \le B$, with $t_T = \lfloor T/2\rfloor$ and $L_T = T - t_T$,
\begin{equation}
  \frac{1}{L_T} \sum_{t=t_T}^{T-1} \E S_{t+1} \le K G^2(1 + \lambda_S) \left( \frac{1}{n} + \frac{1}{T \lambda_S} \right).
  \label{eq:bgf iteration control}
\end{equation}
\end{proposition}

Proposition~\ref{prop:bgf imbalance} bounds the expected score averaged over late selected splits. With fixed regularization, its iteration term is $T^{-1}$ and its sampling term is $n^{-1}$. In Algorithm~\ref{alg:bgf}, this bounds the weighted prognostic discrepancy exposed by late selected trees before the next refit. It quantifies the effect of repeatedly recalibrating the weights on the subsequent split criterion. We next connect this score bound to causal bias under conditions on treatment assignment and the estimated outcome contrasts under $Q$.

\subsection{Threshold recovery}
\label{sec:theory consistency}

We continue under the setup and conditions of Section~\ref{sec:theory feedback}. The final step is a detection argument. If an outcome-relevant treatment-assignment jump remains far from all thresholds selected so far, it creates residual imbalance near that jump. A persistent outcome contrast converts the imbalance into a nonnegligible BGF score. Proposition~\ref{prop:bgf imbalance} bounds the expected scores averaged over late iterations.

\begin{assumption}[Assignment thresholds]
\label{assump:bgf jumps}
For a fixed $J_e \subset [p]$ and fixed finite $L_j$,
\[
  e_1(x) = \alpha_0 + \sum_{j \in J_e} \sum_{\ell=1}^{L_j} c_{j \ell} \one\{x_j > \xi_{j \ell}\},
\]
where $0 < \xi_{j1} < \cdots < \xi_{jL_j} < 1$. Set $J_c = \{j \in J_e:\sum_{a=0}^1 \|f_{aj}\|_2^2 > 0\}$. For $j \in J_c$ and $\ell \le L_j$, $|c_{j \ell}| \ge \kappa$ with fixed $\kappa > 0$.
\end{assumption}

The set $J_c$ contains assignment coordinates that also enter an outcome regression. The product design makes this distinction precise by removing cross-coordinate terms from the residual-confounding covariance.

For the fitted outcome regressions, define
\[
  \widehat{b}_{aj}(u) = Q(\widehat{\mu}_a \mid X_j \le u) - Q(\widehat{\mu}_a \mid X_j > u), \qquad 0<u<1.
\]
We impose the following signal condition on $\widehat{b}_{aj}$ for $j\in J_c$.

\Needspace{6\baselineskip}
\begin{assumption}[Persistent outcome contrast signal]
\label{assump:bgf contrast signal}
There are fixed constants $c_g, \chi > 0$ and $\nu \ge 0$, and a deterministic $\varepsilon_N \in [0, 1]$, such that an event $\mathcal{A}_N \in \sigma(\mathcal{D}_N)$ with $ \Prb(\mathcal{A}_N^c) \le \varepsilon_N$ has the following property. On $\mathcal{A}_N$, for every $j \in J_c$ there is an arm $a(j) \in \{0, 1\}$ such that, for every closed interval $I \subset [0, 1]$ of length $h > 0$, there exists a closed $K \subset I$ with
\[
  \begin{aligned}
    |K| & \ge \chi h \\
    \text{and}\quad \inf_{u \in K \cap (0,1)} |\widehat{b}_{a(j)j}(u)| & \ge c_g h^\nu, 
  \end{aligned}
\]
where $|K|$ denotes the Lebesgue measure of $K$.
\end{assumption}

Let $Z_{j,t}$ be the set of split thresholds selected in coordinate $j$ through iteration $t$, augmented by $0$ and $1$, and define
\[
  d_t = \max_{j \in J_c,\,\ell \le L_j} \min_{z \in Z_{j,t}}|\xi_{j \ell} - z|,
\]
with $d_t = 0$ when $J_c = \varnothing$. Thus $d_t$ is the maximum distance from an outcome-relevant assignment threshold to the corresponding cutoff set $Z_{j,t}$. 

Set
\begin{equation}
  \begin{aligned}
    r_{n,B} & = \frac{p}{m_{\mathrm{try}}}(1 + \lambda_S) \left(\frac{1}{n} + \frac{1}{B \lambda_S}\right) + \frac{B}{n} + \left(\frac{m}{n}\right)^2, \\
    \rho_{n,B} & = \varepsilon_N + r_{n,B}^{1/(\nu+1)}.
  \end{aligned}
  \label{eq:bgf resolution rate}
\end{equation}
The first term comes from late-stage score control and the probability that a relevant coordinate is offered. The term $B/n$ is the empirical cost of the fitted $B$-stump weight class. The minimum-child constraint contributes $(m/n)^2$. The exponent $1/(\nu + 1)$ converts score decay into threshold localization.

\Needspace{7\baselineskip}
\begin{theorem}[ATE consistency]
\label{thm:bgf consistency}
Suppose Assumptions~\ref{assump:theory regularity}, \ref{assump:bgf jumps}, and \ref{assump:bgf contrast signal} hold. For the depth-one BGF construction described above,
\begin{align}
  \E d_B^2&\le K\rho_{n,B}, \\
  \E(\widehat{\theta} - \theta)^2 & \le K \left\{ \frac{M^2 + \sigma_Y^2 + BR^2}{n} + M^2 \rho_{n,B} \right\}. \label{eq:bgf consistency risk}
\end{align}
The constant $K$ does not depend on $(N,n,B,R,M, m, \sigma_Y^2, \lambda_S)$. For fixed $\lambda_S, \lambda_A > 0$, if $B \to \infty,$ $B/n \to 0,$ $m/n \to 0,$ $\varepsilon_N \to 0,$ then $d_B \to 0$ and $\widehat{\theta} \to \theta$ in $L_2$.
\end{theorem}

Theorem~\ref{thm:bgf consistency} proves that the feedback construction resolves the assignment thresholds relevant to outcome bias and that the final calibrated estimator is consistent. An unresolved threshold creates imbalance at a split with a nonnegligible estimated outcome contrast $\widehat{b}_{aj}(u)$. Fresh coordinate offers expose such splits, while Proposition~\ref{prop:bgf imbalance} prevents their scores from remaining large across late iterations. The selected thresholds consequently approach the relevant jumps, and Theorem~\ref{thm:bgf oracle} converts this resolution into ATE risk control. For a single assignment jump and linear outcome regressions, the dependence of residual confounding on cutoff error can be calculated explicitly.

\begin{example}
Let $Q = \mathrm{Unif}(0,1)$, suppose $e_1(x) = p_L + c \one\{x > \xi\}$ satisfies overlap with $0 < \xi < 1$, and let $\mu_a(x) = c_a + \beta_a(x - 1/2)$. Consider a stump with cutoff $u \in (0,1)$. If $u \ne \xi$, only the leaf that contains $\xi$ contributes to residual confounding. Direct integration gives 
\[ 
 P_a|\Delta_{a,1}| = \frac{|c \beta_a|}{2\pi_a} 
  \begin{cases}
  (1 - \xi)(\xi - u), & u \le \xi,\\ 
  \xi(u-\xi),& u \ge \xi.
  \end{cases} 
\] 
The remainder is zero when the stump splits at $\xi$ and is proportional to the cutoff error on either side of the jump.  
\end{example}

For fixed $p, m_{\mathrm{try}}, m, \lambda_S$, choosing $B \asymp n^{1/2}$ balances $B^{-1}$ and $B/n$. With the remaining model constants fixed and negligible $\varepsilon_N$, the displayed mean-squared error is $O(n^{-1/\{2(\nu+1)\}})$. For $\nu = 0$, the rate is $O(n^{-1/2})$, or $O(n^{-1/4})$ root mean-squared error.

\section{Experiments}
\label{sec:experiments}

\lead{Simulation design.}
We set $p = 100$ and generate $X_i \sim N_p(0, \Sigma_\rho)$, where $(\Sigma_\rho)_{jk} = \rho^{|j-k|}$ and $\rho \in \{0, 0.5, 0.8\}$. Let $$S_{ij} = |X_{ij}| - \sqrt{2/\pi}, \qquad A_i \mid X_i \sim \operatorname{Bernoulli}\{e_1(X_i)\},$$ with $e_1(X_i) = 0.1 + 0.8\{1 + \exp[-f(X_i)]\}^{-1}$, and $$f(X_i) = 0.25 \left(\sum_{j=1}^{6} S_{ij} + \sum_{j=21}^{30} S_{ij}\right), \qquad \mu_0(X_i) = 2 \left(\sum_{j=1}^{6} S_{ij} + \sum_{j=11}^{15} S_{ij}\right),$$ $$ Y_i = \mu_0(X_i) + A_i + \varepsilon_i, \qquad \varepsilon_i \sim N(0,1), \qquad \varepsilon_i \perp (X_i,A_i).$$ The true ATE is one. When $\rho = 0$, coordinates 1--6 are confounders, 11--15 are precision variables, and 21--30 are instruments in the usual structural sense.

\begin{remark}[Nonlinear confounding] 
  The Gaussian distribution is centrally symmetric and $e_1(-x)=e_1(x)$, giving $\E(X_j \mid A=1) = \E(X_j \mid A=0) = 0$ for every $j$. Thus the raw covariate means are balanced while the absolute-value features can remain confounded. 
\end{remark}

Each replication uses independent analysis, nuisance, and structure samples of size $n$. Comparators are outcome-guided ($K_Y$) and joint treatment--outcome ($K_J$) forest kernel balancing, RBF kernel balancing, outcome regression (OR), normalized IPW (nIPW), normalized AIPW (nAIPW), and BGF with uniform construction weights (no feedback). BGF, $K_Y$, and $K_J$ share the final ridge-balancing solver. We report RMSE over 100 replications.

\begin{figure}[!tbp]
\centering
\includegraphics[width=\textwidth]{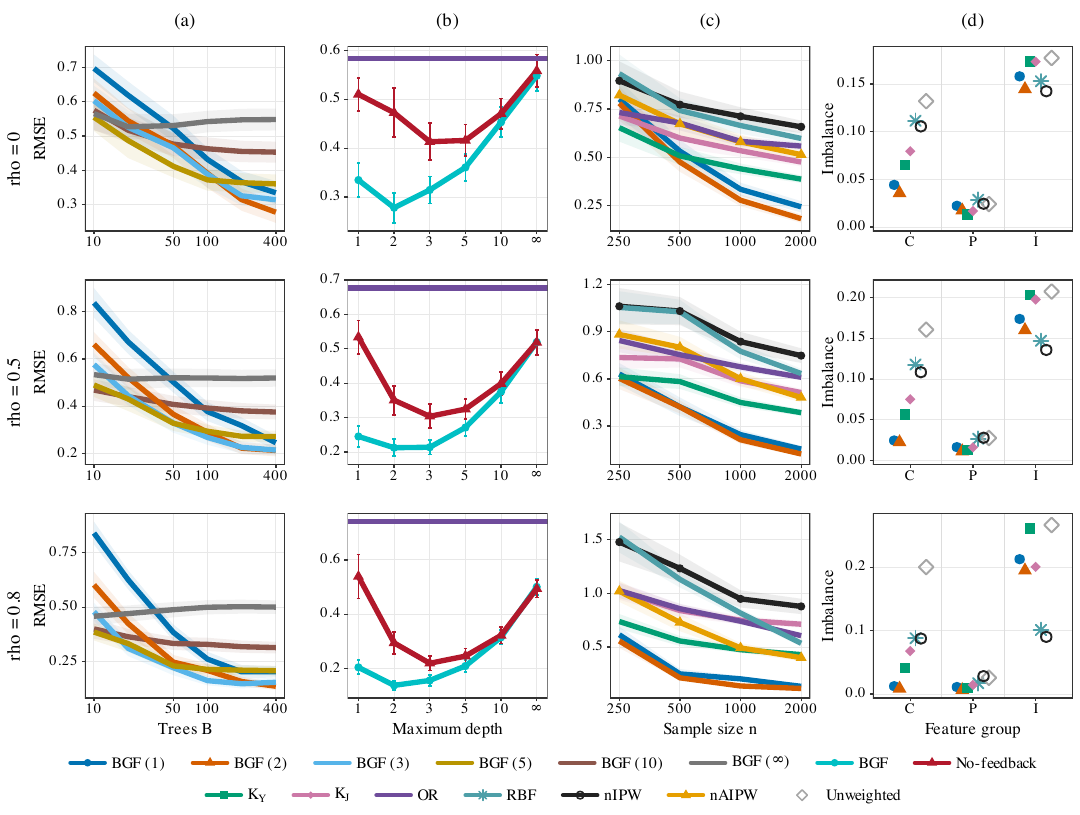}
\caption{Simulations for $\rho = 0, 0.5, 0.8$. ATE RMSE (a) along the BGF path at $n = 1000$; (b) with versus without feedback at $n = 1000$; and (c) versus $n$. (d): mean absolute standardized imbalance for confounding (C), precision (P), and instrumental (I) signals. BGF($d$) has maximum depth $d$ and $d = \infty$ means unrestricted.}
\label{fig:simulation}
\end{figure}

\lead{Feedback is most useful when tree capacity is limited.}
Figure~\ref{fig:simulation}(a) shows that shallow BGF improves steadily as trees are added. At $B = 400$, depth two performs particularly well across all three correlation levels. Deeper trees plateau earlier and at higher RMSE.

Figure~\ref{fig:simulation}(b) isolates feedback. At depths one and two, BGF substantially improves on the no-feedback construction for every $\rho$. The gap narrows with depth and is negligible for unrestricted trees. When a tree can add only a small number of features, each split must be allocated selectively. This is the regime in which feedback should matter most.

\lead{Comparison across sample size.}
Figure~\ref{fig:simulation}(c) compares the shallow BGF estimators with the competing procedures. At $n = 250$, the forest-kernel methods are broadly similar. As $n$ increases, depth-two BGF separates from the alternatives across the three correlation settings.

\lead{Outcome-relevant balance.}
To examine which types of imbalance the methods reduce, we summarize the confounding, precision, and instrumental signals as 
\[
  g_C(X_i) = \sum_{j=1}^{6} S_{ij}, \qquad
  g_P(X_i) = \sum_{j=11}^{15} S_{ij}, \qquad
  g_I(X_i) = \sum_{j=21}^{30} S_{ij}.
\] 

\begin{remark}[Balance and estimation error] 
In this design, $\mu_0=2(g_C+g_P)$ and $\mu_1=\mu_0+1$. Conditional on the constructed forest and the analysis covariates and treatment labels, the weighted estimator in \eqref{eq:bgf ate} satisfies, when both analysis arms are nonempty, 
\[ 
\begin{aligned} 
  &\E\bigl\{(\widehat{\theta} - 1)^2 \mid \mathcal{F}_B, \{(X_i,A_i)\}_{i \in \mathcal{D}_A}\bigr\} \\ 
  &\quad = 4\bigl\{(\widehat{P}_{1,\widehat{w}_1} - \widehat{P}_{0,\widehat{w}_0})(g_C+g_P)\bigr\}^2 + \sum_{a=0}^1 \sum_{i \in \mathcal{D}_A:A_i=a} \widehat{w}_{a,i}^2. 
\end{aligned} 
\] 
Imbalance in the confounding and precision signals enters the outcome-mean difference, while the instrumental signal has no direct contribution to this term. At $\rho=0$, the precision signal is balanced in the population, but its weighted means can differ in a finite sample. Balancing the instrumental signal can also change the weights and hence the noise variance. 
\end{remark}

Figure~\ref{fig:simulation}(d) reports average absolute standardized imbalance in these three signals at $n = 1000$. BGF, especially depth two, gives the smallest confounding-feature imbalance across all $\rho$. Differences for precision features are smaller. BGF does not uniformly minimize instrumental imbalance, as the score in \eqref{eq:ate score} prioritizes imbalance accompanied by outcome separation. Appendix~\ref{app:additional experiments} reports smooth-signal, heterogeneous-effect, tuning-sensitivity, and interaction-dominated experiments.

\section{Real-data applications}
\label{sec:real data}

\lead{Right heart catheterization.}
We analyze the publicly available right heart catheterization (RHC) dataset from the study of \citet{connors1996}. The sample contains 5,735 critically ill patients, of whom 2,184 received RHC within 24 hours of admission. Following \citet{de2025}, we estimate the target-population risk difference for 30-day mortality.

\lead{Smoking cessation.}
We also analyze the National Health and Nutrition Examination Survey Epidemiologic Follow-up Study (NHEFS) \citep{nchs1987}. Following \citet{ertefaie2023}, we use the public analysis data of \citet{hernan2020}, with 1,566 individuals. Treatment is smoking cessation and the outcome is subsequent weight gain in kilograms.

For all methods except EmpDiff, we use the same tuning settings as in the simulations and repeat cross-fitting five times, dividing the sample into three folds stratified by treatment status in each repetition. Appendix~\ref{app:real data implementation} gives data and implementation details.

\begin{table}[!htbp]
\centering
\caption{ATE estimates in real-data applications. }
\label{tab:real data ate}
\setlength{\tabcolsep}{10pt}
\begin{tabular}{lrr}
  \toprule
  Method & RHC (pp) & NHEFS (kg) \\
  \midrule
  EmpDiff & 7.364 & 2.541 \\
  BGF(1)  & 4.553 & 3.440 \\
  BGF(2)  & 5.021 & 3.413 \\
  $K_Y$   & 4.978 & 3.420 \\
  $K_J$   & 5.250 & 3.238 \\
  OR      & 5.321 & 3.334 \\
  RBF     & 5.824 & 3.627 \\
  nIPW    & 3.789 & 3.550 \\
  nAIPW   & 4.985 & 3.570 \\
  \bottomrule
\end{tabular}
\end{table}

Table~\ref{tab:real data ate} shows that the adjusted procedures produce broadly similar estimates in both studies and differ from the empirical difference in means. The adjusted point estimates are positive for RHC mortality and for weight gain after smoking cessation.

\begin{figure}[!tbp]
\centering
\includegraphics[width=0.90\linewidth]{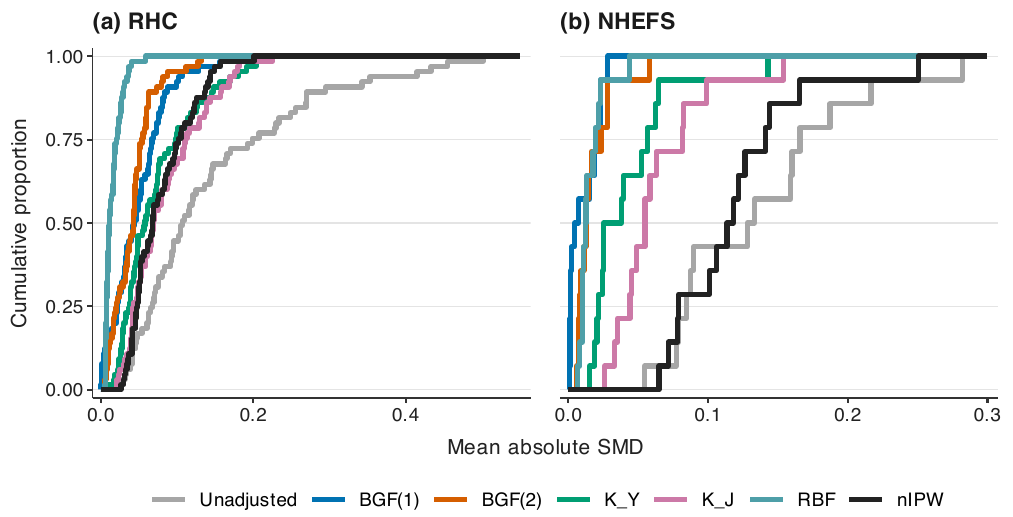}
\caption{Covariate-balance ECDFs for RHC and NHEFS. Upper-left curves indicate better balance.}
\label{fig:real data balance}
\end{figure}

Figure~\ref{fig:real data balance} summarizes marginal balance using weighted standardized mean differences \citep{austin2015}. In RHC, RBF attains the smallest marginal imbalances, while the imbalances under both BGF methods are concentrated near zero relative to the other forest-kernel and weighting methods shown. In NHEFS, BGF and RBF give the strongest marginal balance. BGF is designed to prioritize outcome-relevant imbalance, so it need not minimize marginal covariate imbalance.

\FloatBarrier
\section{Discussion}

BGF uses current weighting error to decide what the forest should represent next. Shallow trees allow more frequent recalibration per split. The weights are refitted between trees but held fixed within a tree. After the root split, deeper splits are therefore evaluated using weights attached to a representation that has already changed. This offers one possible explanation for the stronger empirical gains at small depth.

Depth also controls interaction capacity. A stump receives no marginal guidance toward a pure interaction with zero root-level outcome contrast, as shown in Appendix~\ref{app:pure interaction stress test}. Deeper trees can reveal such structure, but they weaken synchronization between the current representation and the current weights. Updating weights within tree growth is a natural extension, as it would preserve the bias-directed score while allowing richer interaction structure.

\section*{Acknowledgments}
\phantomsection\label{sec:acknowledgments}

Generative AI tools were used in manuscript preparation and for assistance with programming, data processing, and exploring directions in formal proofs. All theoretical arguments, implementations, and numerical results were independently verified by the authors, who take full responsibility for the final work.

\FloatBarrier
\bibliographystyle{abbrvnat}
\bibliography{reference}

\FloatBarrier
\appendix
\Needspace{8\baselineskip}
\section*{Appendices}
\phantomsection
\addcontentsline{toc}{section}{Appendices}

\section{Proofs for Section~\ref{sec:theory}}
\label{app:theory proofs}

Throughout the appendix, constants denoted by $K$ and $c$ may change from line to line. They depend only on the fixed constants stated in the corresponding result. For a probability measure $P$, write $\|f\|_P^2 = Pf^2$. Empty-arm empirical averages are set to zero. Quantities introduced only on a specified event are extended by zero elsewhere, unless another convention is stated. Equations involving them are restricted to that event.

\subsection{Proof of Theorem~\ref{thm:bgf oracle}}
\label{app:proof oracle}

Write $Q_A$ for the empirical covariate law on $\mathcal{D}_A$, $P_{A,a}$ for its empirical law within arm $a$, and $n_{A,a}$ for the arm count. The empirical averages also apply to functions involving $Y$.

Set $\widehat{\theta} = 0$ if an analysis arm is empty. When both analysis arms are nonempty, $\widehat{v}_a = n_{A,a} \widehat{w}_a$ solves
\begin{equation}
  \label{eq:theory analysis refit}
  \widehat{v}_a = \argmin_{\substack{0 \le v \le C\\ P_{A,a} v=1}} \left\{ \|P_{A,a}(v \Phi_B) - Q_A \Phi_B\|^2 + \frac{\lambda_A}{n_{A,a}} P_{A,a} v^2 \right\},
\end{equation}
and
\[
  \widehat{\theta} = P_{A,1}(\widehat{v}_1 Y) - P_{A,0}(\widehat{v}_0 Y).
\]

\begin{proof}
Condition on the realized construction $\mathcal{F}_B$ and write $\mathscr{F} = \sigma\{\Phi_B(X)\}$. Since $dP_a/dQ = e_a/\pi_a$, for every integrable $h$,
\begin{equation}
  \label{eq:app change measure}
  \E_{P_a}(h \mid \mathscr{F}) = \frac{\E_Q(e_a h \mid \mathscr{F})}{\E_Q(e_a \mid \mathscr{F})}.
\end{equation}
Assumption~\ref{assump:theory regularity} implies $e_a \ge \pi_a/C_*$, so the denominator is bounded away from zero. Applying \eqref{eq:app change measure} to $h = \mu_a$ and using the definitions of $\mu_{a,B}^Q$, $\mu_{a,B}^P$, and $e_{a,B}^Q$ gives
\begin{equation}
  \label{eq:app covariance identity}
  e_{a,B}^Q \Delta_{a,B} = \E_Q \{(e_a - e_{a,B}^Q)(\mu_a - \mu_{a,B}^Q) \mid \mathscr{F}\}.
\end{equation}
Because $\Delta_{a,B}$ is $\mathscr{F}$-measurable,
\begin{align}
  P_a|\Delta_{a,B}| & = \pi_a^{-1} Q(e_a|\Delta_{a,B}|) = \pi_a^{-1} Q(e_{a,B}^Q|\Delta_{a,B}|)\nonumber\\
  & \le \pi_a^{-1} Q|(e_a - e_{a,B}^Q)(\mu_a - \mu_{a,B}^Q)|\nonumber\\
  & \le \pi_a^{-1}\|e_a - e_{a,B}^Q\|_Q\|\mu_a - \mu_{a,B}^Q\|_Q,
  \label{eq:app product bound}
\end{align}
which proves \eqref{eq:bgf product remainder}.

We next represent $\mu_{a,B}^Q$ in the actual stump span. For coordinate $j$, let
\[
  0 = z_{j,0} < z_{j,1} < \cdots < z_{j,K_j} < z_{j,K_j+1} = 1
\]
be the distinct split thresholds in coordinate $j$, together with the endpoints, and define $I_{j,r} = (z_{j,r-1}, z_{j,r}]$. Let $\Pi_j f$ be the average of $f$ over the interval $I_{j,r}$ containing $X_j$. Since $Q$ is product measure and the outcome regression is additive,
\begin{equation}
  \mu_{a,B}^Q = c_a + \sum_{j=1}^p \Pi_j f_{aj}, \qquad Q \mu_{a,B}^Q = Q \mu_a.
\end{equation}
Write
\[
  c_{a,j,r} = |I_{j,r}|^{-1} \int_{I_{j,r}} f_{aj}(u)\,du, \qquad \delta_{a,j,r} = c_{a,j,r} - c_{a,j,r+1}.
\]
For adjacent intervals $I = [u_0, u_1]$ and $J = [u_1, u_2]$, with lengths $h_I$ and $h_J$, integration by parts around the common endpoint gives
\[
\frac{1}{h_I} \int_I f_{aj} - \frac{1}{h_J} \int_J f_{aj} = -\int_{u_0}^{u_1} \frac{s - u_0}{h_I} f'_{aj}(s)\,ds - \int_{u_1}^{u_2} \frac{u_2 - s}{h_J} f'_{aj}(s)\,ds.
\]
Cauchy--Schwarz therefore yields the slightly sharper bound
\[
  \left(\frac{1}{h_I} \int_I f_{aj} - \frac{1}{h_J} \int_J f_{aj}\right)^2 \le \frac{h_I + h_J}{3} \int_{I \cup J} \{f'_{aj}(s)\}^2 ds,
\]
and in particular
\begin{equation}
  \delta_{a,j,r}^2 \le \{|I_{j,r}| + |I_{j,r+1}|\} \int_{I_{j,r} \cup I_{j,r+1}} \{f'_{aj}(s)\}^2 ds.
\end{equation}
Each interval appears in at most two neighboring pairs and each adjacent length sum is at most two. Hence, 
\begin{equation}
  \sum_{a,j,r} \delta_{a,j,r}^2 \le 4 R^2.
\end{equation}
Moreover,
\begin{equation}
  \label{eq:app telescope}
  \Pi_j f_{aj} = c_{a,j,K_j+1} + \sum_{r=1}^{K_j} \delta_{a,j,r} \one\{X_j \le z_{j,r}\}.
\end{equation}
Assign each distinct threshold in \eqref{eq:app telescope} to its first occurrence among the $B$ stumps. Since each stump contributes the two indicators $(\one\{X_j \le z\}, \one\{X_j > z\})$ scaled by $B^{-1/2}$, the constants can be absorbed into an intercept and there is a vector $\gamma_a$ such that
\begin{equation}
  \label{eq:app outcome representation}
  \mu_{a,B}^Q = c_{a,B} + \gamma_a^\top \Phi_B, \qquad \sum_{a=0}^1 \|\gamma_a\|^2 = \frac{B}{2} \sum_{a,j,r} \delta_{a,j,r}^2 \le 2BR^2.
\end{equation}

We now control the analysis-sample feature imbalance. Let
\[
  \mathcal{N}_A = \{n_{A,0} n_{A,1} > 0\}
\]
and define the event
\[
  \mathcal{G}_A = \bigcap_{a=0}^1 \left\{ n_{A,a} \ge n \pi_*/2, \ |P_{A,a} w_a^\star - 1| \le (C - C_*)/(2C) \right\}.
\]

Because $n_{A,a} \sim \mathrm{Binomial}(n, \pi_a)$ and $\pi_a \ge \pi_*$, Hoeffding's inequality gives $\Prb\{n_{A,a} < n \pi_*/2\} \le e^{-cn}$. Conditional on $n_{A,a} = q > 0$, the arm-$a$ covariates are iid from $P_a$, and $0 \le w_a^\star \le C_*$ with $P_a w_a^\star = 1$. A second application of Hoeffding therefore gives
\[
  \Prb \left\{|P_{A,a} w_a^\star - 1| > (C - C_*)/(2C), \ n_{A,a} \ge n \pi_*/2\right\} \le Ke^{-cn}.
\]
A union bound yields $\Prb(\mathcal{G}_A^c \mid \mathcal{F}_B) \le Ke^{-cn}$. For $K_a \sim \mathrm{Binomial}(n, \pi_a)$,
\[
  \E \frac{1}{K_a + 1} = \int_0^1 (1 - \pi_a + \pi_a t)^n\,dt = \frac{1 - (1 - \pi_a)^{n+1}}{(n + 1) \pi_a}.
\]
Since $\one(k > 0)/k \le 2/(k + 1)$,
\begin{equation}
  \label{eq:app inverse count}
  \E\{\one(n_{A,a} > 0)/n_{A,a} \mid \mathcal{F}_B\} \le \frac{2}{(n + 1) \pi_a} \le \frac{2}{n \pi_*}.
\end{equation}

On $\mathcal{G}_A$, the normalized inverse-probability comparator
\[
  v_{A,a}^\star = \frac{w_a^\star}{P_{A,a} w_a^\star}
\]
is feasible for \eqref{eq:theory analysis refit}, satisfies $0 \le v_{A,a}^\star < C$, and obeys $P_{A,a}(v_{A,a}^\star)^2 \le C$. Define 
\[
  \widehat{r}_a = P_{A,a}(\widehat{v}_a \Phi_B) - Q_A \Phi_B, \qquad r_a^\star = P_{A,a}(v_{A,a}^\star \Phi_B) - Q_A \Phi_B.
\]
Using $P_a(w_a^\star \Phi_B) = Q \Phi_B$,
\begin{equation}
  \label{eq:app oracle analysis residual}
  r_a^\star = \frac{(P_{A,a} - P_a)(w_a^\star \Phi_B)}{P_{A,a} w_a^\star} + \{(P_{A,a} w_a^\star)^{-1} - 1\}Q \Phi_B - (Q_A - Q) \Phi_B.
\end{equation}

We bound the three terms in \eqref{eq:app oracle analysis residual} separately. Conditional on $n_{A,a} = q > 0$, the $q$ covariates in arm $a$ are iid from $P_a$, so
\[
  \E \left[\left\|(P_{A,a} - P_a)(w_a^\star \Phi_B)\right\|^2 \Bigm|\mathcal{F}_B, n_{A,a} = q\right] \le \frac{C_*^2}{q},
\]
and similarly
\[
  \E \left[|P_{A,a} w_a^\star - 1|^2 \Bigm|\mathcal{F}_B, n_{A,a} = q\right] \le \frac{C_*^2}{q}.
\]
The marginal covariates in the full analysis sample are iid from $Q$, independently of $\mathcal{F}_B$, and $\|\Phi_B(X)\| = 1$, hence
\[
  \E\{\|(Q_A - Q) \Phi_B\|^2 \mid \mathcal{F}_B\} \le \frac{1}{n}.
\]

On $\mathcal{G}_A$, $P_{A,a} w_a^\star$ is bounded below by $1 - (C - C_*)/(2C) > 1/2$. Thus $(x + y + z)^2 \le 3(x^2 + y^2 + z^2)$, followed by \eqref{eq:app inverse count}, gives
\begin{equation}
  \label{eq:app oracle analysis balance}
  \E\{\one_{\mathcal{G}_A}\|r_a^\star\|^2 \mid \mathcal{F}_B\} \le K/n.
\end{equation}
Optimality of $\widehat{v}_a$ against $v_{A,a}^\star$ implies
\[
  \|\widehat{r}_a\|^2 + \frac{\lambda_A}{n_{A,a}} P_{A,a} \widehat{v}_a^2 \le \|r_a^\star\|^2 + \frac{\lambda_A}{n_{A,a}} P_{A,a}(v_{A,a}^\star)^2.
\]
Drop the nonnegative ridge term on the left and use $P_{A,a}(v_{A,a}^\star)^2 \le C$ to obtain
\[
  \|\widehat{r}_a\|^2 \le \|r_a^\star\|^2 + \frac{C \lambda_A}{n_{A,a}}.
\]
Thus \eqref{eq:app oracle analysis balance} and \eqref{eq:app inverse count} give an $O(n^{-1})$ contribution on $\mathcal{G}_A$. On $\mathcal{N}_A \cap \mathcal{G}_A^c$, the normalization and cap imply
\[
  \|P_{A,a}(\widehat{v}_a \Phi_B)\| \le P_{A,a}(\widehat{v}_a\|\Phi_B\|) = 1, \qquad \|Q_A \Phi_B\| \le 1,
\]
so $\|\widehat{r}_a\| \le 2$. Since $\Prb(\mathcal{G}_A^c \mid \mathcal{F}_B) \le Ke^{-cn}$,
\begin{equation}
  \label{eq:app analysis balance}
  \E\{\one_{\mathcal{N}_A}\|\widehat{r}_a\|^2 \mid \mathcal{F}_B\} \le K/n.
\end{equation}

The strictly convex analysis program has a unique minimizer measurable in $\mathcal{F}_B$, the analysis labels, and the analysis feature vectors. Conditional on these quantities, the observations remain independent and the weights are fixed. In particular,
\[
  \E\{Y\mid A=a,\Phi_B(X)\}=\mu_{a,B}^P. 
\] 
Moreover, 
\[ 
  \Var\{Y\mid A=a,\Phi_B(X)\}\le \sigma_Y^2+M^2.
\]

Define $Z_a = P_{A,a} \{\widehat{v}_a(Y - \mu_{a,B}^P)\}$ on $\mathcal{N}_A$ and zero otherwise. On $\mathcal{N}_A$,
\begin{equation}
  \label{eq:app outcome noise}
  \begin{aligned}
    \E(Z_a \mid \mathcal{F}_B, A_{1:n}, \Phi_B(X_{1:n})) & = 0, \\
    \E(Z_a^2 \mid \mathcal{F}_B, A_{1:n}, \Phi_B(X_{1:n})) & \le \frac{C(M^2 + \sigma_Y^2)}{n_{A,a}},
  \end{aligned}
\end{equation}
and \eqref{eq:app outcome representation} gives the exact decomposition
\begin{equation}
  \label{eq:app oracle decomposition}
  P_{A,a}(\widehat{v}_a Y) - Q \mu_a = \gamma_a^\top \widehat{r}_a + (Q_A - Q) \mu_{a,B}^Q + P_{A,a}(\widehat{v}_a \Delta_{a,B}) + Z_a.
\end{equation}
The first term has conditional second moment at most $K\|\gamma_a\|^2/n$ by \eqref{eq:app analysis balance}. The second has second moment at most $M^2/n$. For the third term, cappedness gives
\[
  |P_{A,a}(\widehat{v}_a \Delta_{a,B})| \le C P_{A,a}|\Delta_{a,B}|,
\]
and, conditional on the treatment labels with $n_{A,a} > 0$, 
\begin{equation}
  \label{eq:app delta second}
  \E\{(P_{A,a}|\Delta_{a,B}|)^2 \mid \mathcal{F}_B, A_{1:n}\} \le (P_a|\Delta_{a,B}|)^2 + \frac{4 M^2}{n_{A,a}}.
\end{equation}

Apply $(x_1+\cdots+x_4)^2 \le 4\sum_{r=1}^4 x_r^2$ to \eqref{eq:app oracle decomposition} and sum over the two arms. Bound the resulting terms using \eqref{eq:app inverse count}, \eqref{eq:app outcome representation}, \eqref{eq:app analysis balance}, \eqref{eq:app outcome noise}, and \eqref{eq:app delta second}.

On $\mathcal{N}_A^c$, which has exponentially small probability, use $|\widehat{\theta} - \theta| \le 2M$ under the convention stated above. Combining these bounds gives 
\[
  \E\{(\widehat{\theta} - \theta)^2 \mid \mathcal{F}_B\} \le K \left\{ \frac{M^2 + \sigma_Y^2}{n} + \frac{BR^2}{n} + \sum_a (P_a|\Delta_{a,B}|)^2 \right\},
\]
which is \eqref{eq:bgf oracle risk}. The product form follows from \eqref{eq:app product bound}.
\end{proof}

\subsection{Proof of Lemma~\ref{lem:bgf descent}}
\label{app:proof descent}

We use the depth-one construction specified in Section~\ref{sec:theory feedback}. If either structure arm is empty, we bypass structure fitting, use root-only stumps, and set all selected scores to zero. For Lemma~\ref{lem:bgf descent}, we also set $\mathcal{Q}_t(v_t) = 2t\lambda_S/n$ and set the residual and weight-increment terms on the right-hand side of \eqref{eq:bgf descent} to zero. The calculations below concern the case in which both structure arms are nonempty.

If $w_{a,i}^{(t)}$ are the construction weights from \eqref{eq:ate refit}, define the mean-one rescaling
\[
  v_{a,t}(X_i) = n_{S,a} w_{a,i}^{(t)}, \qquad i \in \mathcal{D}_S, \ A_i = a.
\]
Then $P_{S,a} v_{a,t} = 1$ and $0 \le v_{a,t} \le C$. If the split selected at iteration $b$ is $(j_b, u_b)$, define 
\[
  \phi_b(x) = \big(\one\{x_{j_b} \le u_b\}, \one\{x_{j_b} > u_b\}\big)^\top, \qquad \Phi_t = t^{-1/2}(\phi_1^\top, \ldots, \phi_t^\top)^\top.
\]
The construction refit is equivalently
\begin{equation}
  \label{eq:theory structure refit}
  v_{a,t} = \argmin_{\substack{0 \le v \le C\\ P_{S,a} v=1}} \left\{ \|P_{S,a}(v \Phi_t) - Q_S \Phi_t\|^2 + \frac{\lambda_S}{n_{S,a}} P_{S,a} v^2 \right\}.
\end{equation}
The identities above are simply reparameterizations of \eqref{eq:ate refit}: since $v_{a,t} = n_{S,a} w_a^{(t)}$, the ridge term $\lambda_S\|w_a^{(t)}\|_2^2$ equals $(\lambda_S/n_{S,a})P_{S,a} v_{a,t}^2$.

\begin{proof}
For one arm, the scaling of $\Phi_t$ gives
\[
  \|P_{S,a}(v \Phi_t) - Q_S \Phi_t\|^2 = \frac{1}{t} \sum_{b=1}^t \|R_{a,b}(v)\|^2.
\]
Multiplying the arm-$a$ criterion in \eqref{eq:theory structure refit} by the positive constant $t n_{S,a}/n$ shows that the pair $v_t = (v_{0,t}, v_{1,t})$ minimizes $\mathcal{Q}_t$ in \eqref{eq:feedback criterion} over the product of the two normalized boxes.

Fix a feasible $u = (u_0, u_1)$ and write $h = u - v_t$. Since $\mathcal{Q}_t$ is quadratic,
\begin{align}
  \mathcal{Q}_t(v_t + h) - \mathcal{Q}_t(v_t) = {} & D \mathcal{Q}_t(v_t)[h] + \frac{t \lambda_S}{n} \sum_a P_{S,a} h_a^2 + \sum_a \frac{n_{S,a}}{n} \sum_{b=1}^t \|P_{S,a}(h_a \phi_b)\|^2.
\end{align}
The feasible set is convex. Hence $v_t + sh$ is feasible for $0 \le s \le 1$, and the one-sided derivative at the minimizer is nonnegative. Therefore
\begin{equation}
  \label{eq:app strong convexity}
  \mathcal{Q}_t(u) - \mathcal{Q}_t(v_t) \ge \frac{t \lambda_S}{n} \sum_a \|u_a - v_{a,t}\|_{P_{S,a}}^2.
\end{equation}

For fixed $u$, increasing the iteration index adds one penalty increment and the residual of the new stump:
\begin{equation}
  \label{eq:app criterion increment}
  \mathcal{Q}_{t+1}(u) - \mathcal{Q}_t(u) = \frac{\lambda_S}{n} \sum_a P_{S,a} u_a^2 + \sum_a \frac{n_{S,a}}{n}\|R_{a,t+1}(u_a)\|^2.
\end{equation}
Apply \eqref{eq:app strong convexity} with $u = v_{t+1}$ and add \eqref{eq:app criterion increment}. Since $P_{S,a} v_{a,t+1} = 1$, Jensen's inequality gives $P_{S,a} v_{a,t+1}^2 \ge 1$ for each arm. Subtracting $2 \lambda_S/n$ proves \eqref{eq:bgf descent}.
\end{proof}

\subsection{Proof of Proposition~\ref{prop:bgf imbalance}}
\label{app:proof imbalance}

For coordinate $j$, let $\mathcal{U}_j$ be the root cutoffs between consecutive ordered structure observations that leave at least $m$ observations in each child. Under Assumption~\ref{assump:theory regularity}, coordinate ties occur with probability zero.

For $H_{j,u} = \{x:x_j \le u\}$, define at construction iteration $t$
\[
  I_{a,t}(j, u) = P_{S,a} \{v_{a,t} \one_{H_{j,u}}\} - Q_S H_{j,u},
\]
\[
  H_{a}(j, u) = Q_S(\widehat{\mu}_a \mid H_{j,u}) - Q_S(\widehat{\mu}_a \mid H_{j,u}^c).
\]
At a root split, \eqref{eq:ate score} is exactly
\begin{equation}
  \label{eq:theory score}
  \Gamma_{t+1}(j, u) = \sum_{a=0}^1 \{I_{a,t}(j, u)H_a(j, u)\}^2.
\end{equation}
Let $(j_{t+1}, u_{t+1})$ maximize \eqref{eq:theory score} over the offered coordinates and admissible root cutoffs, and define
\[
  S_{t+1} = \Gamma_{t+1}(j_{t+1}, u_{t+1}).
\]

\begin{proof}
Fix $2 \le T \le B$, define $t_T = \lfloor T/2\rfloor$ and $L_T = T - t_T$, and let
\[
  \mathscr{H} = \{\{x:x_j \le u\}, \{x:x_j > u\}: j \in [p], \ u \in [0, 1]\} \cup \{[0, 1]^p\}.
\]
The constants below are uniform in $T$.

\emph{Step 1: an oracle upper bound for the terminal quadratic criterion.} Let
\[
  \mathcal{C}_n = \{\min_a n_{S,a} \ge n \pi_*/2\}
\]
and
\[
  \mathcal{C}_w = \mathcal{C}_n \cap \bigcap_a \left\{|P_{S,a} w_a^\star - 1| \le (C - C_*)/(2C)\right\}.
\]
On $\mathcal{C}_w$, define
\[
  v_{S,a}^\star = \frac{w_a^\star}{P_{S,a} w_a^\star}.
\]
Then $P_{S,a} v_{S,a}^\star = 1$ and
\[
  0 \le v_{S,a}^\star \le \frac{2CC_*}{C + C_*} < C, \qquad P_{S,a}(v_{S,a}^\star)^2 \le C.
\]
Set
\[
  \delta_S = \max_a \sup_{H \in \mathscr{H}} |P_{S,a}(v_{S,a}^\star \one_H) - Q_S H|.
\]
Writing $m_a = P_{S,a} w_a^\star$, the population identity $P_a(w_a^\star \one_H) = QH$ gives
\[
  P_{S,a}(v_{S,a}^\star \one_H) - Q_S H = \frac{(P_{S,a} - P_a)(w_a^\star \one_H) - (m_a - 1)QH}{m_a} + (Q - Q_S)H.
\]
On $\mathcal{C}_w$, $m_a$ is bounded away from zero. For a stump with left leaf $H$, normalization then makes the two coordinates of $R_{a,b}(v_{S,a}^\star)$ equal to $(d, -d)$ with $|d| \le \delta_S$, so $\|R_{a,b}(v_{S,a}^\star)\|^2 \le 2 \delta_S^2$. By optimality of $v_T$,
\begin{equation}
  \label{eq:app terminal criterion}
  \mathcal{Q}_T(v_T) \le \mathcal{Q}_T \big((v_{S,0}^\star, v_{S,1}^\star) \big) \le \frac{2CT \lambda_S}{n} + 2T \delta_S^2 \qquad \text{on } \mathcal{C}_w.
\end{equation}

We give a fixed-function bound that will also be applied to $\widehat{\mu}_a$ after conditioning on its independent training sample $\mathcal D_N$. Let $X_1, \ldots, X_q$ be iid from a law $P$, let $P_q$ be their empirical law, and let $h$ be a fixed measurable function with $|h| \le U$. Introduce an independent copy of the sample and independent Rademacher signs $\varepsilon_i$. Jensen's inequality, sample--copy exchange, and $(x + y)^2 \le 2x^2 + 2y^2$ give
\[
  \E \sup_{H \in \mathscr{H}} |(P_q - P)(h \one_H)|^2 \le 4 \E \sup_{H \in \mathscr{H}} \left|q^{-1} \sum_{i=1}^q \varepsilon_i h(X_i) \one_H(X_i)\right|^2.
\]
Conditional on the covariates, order observations along one coordinate. The signed partial sums of $h(X_i)$ form a martingale with terminal second moment $\sum_i h(X_i)^2 \le q U^2$. Its expected squared maximum is at most $4q U^2$ by the $L_2$ maximal inequality. The $p$ increasing and $p$ decreasing orders cover all members of $\mathscr{H}$, including the full cube. Bounding a maximum by a sum therefore proves 
\begin{equation}
  \label{eq:bounded halfspace moment}
  \E \sup_{H \in \mathscr{H}} |(P_q - P)(h \one_H)|^2 \le \frac{32p U^2}{q}.
\end{equation}
The bound applies to signed $h$ as well. We apply this bound to the fixed function $w_a^\star$ here and later to $\widehat{\mu}_a$ after conditioning on its independent training sample.

Conditional on $n_{S,a} = q > 0$, use $P = P_a$ and $h = w_a^\star$ in \eqref{eq:bounded halfspace moment}. Use $P = Q$ and $h = 1$ on the full structure sample. This gives
\[
  \E \left\{\sup_{H \in \mathscr{H}} |(P_{S,a} - P_a)(w_a^\star \one_H)|^2 \mid n_{S,a} = q\right\} \le \frac{32pC_*^2}{q}, \qquad \E \sup_{H \in \mathscr{H}} |(Q_S - Q)H|^2 \le \frac{32p}{n}.
\]

On $\mathcal{C}_n$, $n_{S,a} \ge n \pi_*/2$ by definition, and therefore
\begin{equation}
  \E \left[ \one_{\mathcal{C}_n} \left\{ \sup_{H \in \mathscr{H}} |(Q_S - Q)H|^2 + \max_a \sup_{H \in \mathscr{H}} |(P_{S,a} - P_a)(w_a^\star \one_H)|^2 \right\} \right] \le K/n.
\end{equation}
Since the full cube belongs to $\mathscr{H}$, the same bound controls $|P_{S,a} w_a^\star - 1|$. The displayed identity for $P_{S,a}(v_{S,a}^\star \one_H) - Q_S H$, together with the lower bound on $m_a$ on $\mathcal{C}_w$, therefore gives
\begin{equation}
  \label{eq:app deltaS second moment}
  \E(\one_{\mathcal{C}_w} \delta_S^2) \le K/n.
\end{equation}
The binomial lower-tail bound for $n_{S,a}$ and Hoeffding's inequality for the bounded arm average $P_{S,a} w_a^\star$ imply $\Prb(\mathcal{C}_w^c) \le Ke^{-cn}$. On the nonbypass part of $\mathcal{C}_w^c$, the constant vector one is feasible and yields $\mathcal{Q}_T(v_T) \le KT(1 + \lambda_S/n)$. On a bypass run, the convention in Appendix~\ref{app:proof descent} gives the smaller value $2T \lambda_S/n$. Combining these facts with \eqref{eq:app terminal criterion}--\eqref{eq:app deltaS second moment} gives
\begin{equation}
  \label{eq:app mean terminal criterion}
  \E \mathcal{Q}_T(v_T) \le KT \left(\frac{\lambda_S}{n} + \frac{1}{n}\right).
\end{equation}

\emph{Step 2: control the pre-refit score.} Let $H = H_{j_{t+1},u_{t+1}}$ be the selected left leaf. Define 
\[
  d_a = P_{S,a}(v_{a,t} \one_H) - Q_S H, \qquad d_a' = P_{S,a}(v_{a,t+1} \one_H) - Q_S H.
\]
Then
\[
  d_a = d_a' - P_{S,a} \{(v_{a,t+1} - v_{a,t}) \one_H\}.
\]
Since $P_{S,a} v_{a,t+1} = 1$, the residual on $H^c$ is $-d_a'$, and therefore
\begin{equation}
  \label{eq:app new stump residual}
  \|R_{a,t+1}(v_{a,t+1})\|^2 = 2(d_a')^2.
\end{equation}
Also,
\begin{equation}
  \label{eq:app halfspace weight change}
  \left|P_{S,a} \{(v_{a,t+1} - v_{a,t}) \one_H\}\right|^2 \le \|v_{a,t+1} - v_{a,t}\|_{P_{S,a}}^2.
\end{equation}
The $\widehat{\mu}_a$ contrast in \eqref{eq:theory score} is bounded by $2G$. Thus $(x + y)^2 \le 2x^2 + 2y^2$, together with \eqref{eq:app new stump residual}--\eqref{eq:app halfspace weight change}, gives
\begin{equation}
  \label{eq:app score control}
  S_{t+1} \le 4G^2 \sum_a \|R_{a,t+1}(v_{a,t+1})\|^2 + 8G^2 \sum_a \|v_{a,t+1} - v_{a,t}\|_{P_{S,a}}^2.
\end{equation}

\emph{Step 3: sum the descent inequality.} Let $D_t = \mathcal{Q}_{t+1}(v_{t+1}) - \mathcal{Q}_t(v_t) - 2 \lambda_S/n$. By Lemma~\ref{lem:bgf descent}, $D_t$ dominates both nonnegative terms on the right of \eqref{eq:bgf descent}. Moreover, Jensen's inequality gives $\mathcal{Q}_{t_T}(v_{t_T}) \ge 2t_T \lambda_S/n$, so
\begin{align}
  \sum_{t=t_T}^{T-1} D_t & = \mathcal{Q}_T(v_T) - \mathcal{Q}_{t_T}(v_{t_T}) - \frac{2L_T \lambda_S}{n}\nonumber\\
  & \le \mathcal{Q}_T(v_T) - \frac{2T \lambda_S}{n} \le \mathcal{Q}_T(v_T).
  \label{eq:app descent budget}
\end{align}

On $\mathcal{C}_n$, note that $n_{S,a}/n \ge \pi_*/2$. Hence \eqref{eq:app mean terminal criterion} and \eqref{eq:app descent budget} imply
\begin{align}
  \E \left[ \one_{\mathcal{C}_n} \sum_{t=t_T}^{T-1} \sum_a \|R_{a,t+1}(v_{a,t+1})\|^2 \right] & \le KT \left(\frac{\lambda_S}{n} + \frac{1}{n}\right),
  \label{eq:app residual budget}
  \\
  \E \sum_{t=t_T}^{T-1} \sum_a \|v_{a,t+1} - v_{a,t}\|_{P_{S,a}}^2 & \le K \left(1 + \frac{1}{\lambda_S}\right).
  \label{eq:app weight movement budget}
\end{align}
For the second line, use $t \ge t_T \ge T/3$ for $T \ge 2$ in the first term of \eqref{eq:bgf descent}. Sum \eqref{eq:app score control} and apply \eqref{eq:app residual budget}--\eqref{eq:app weight movement budget}. On $\mathcal{C}_n^c$, $0 \le P_{S,a}(v_{a,t} \one_H), Q_S H \le 1$ and $|H_a(j, u)| \le 2G$, so deterministically $S_{t+1} \le 8G^2$. The probability of $\mathcal{C}_n^c$ is exponentially small. Since $L_T \ge T/2$, we have
\[
  \frac{1}{L_T} \sum_{t=t_T}^{T-1} \E S_{t+1} \le KG^2 \left\{ \frac{1 + \lambda_S}{n} + \frac{1 + \lambda_S}{T \lambda_S} \right\},
\]
which is \eqref{eq:bgf iteration control}.
\end{proof}

\subsection{Proof of Theorem~\ref{thm:bgf consistency}}
\label{app:proof consistency}

Use the depth-one construction described in Section~\ref{sec:theory feedback} and Appendix~\ref{app:proof descent}. Recall that $\mathcal{C}_n = \{\min_a n_{S,a} \ge n \pi_*/2\}$. Let $\mathcal{M}_t$ be the set offered at iteration $t$ and define 
\[
  \mathcal{F}_t = \sigma(\mathcal{D}_N, \mathcal{D}_S, \mathcal{M}_1, \ldots, \mathcal{M}_t), \qquad \mathcal{F}_0 = \sigma(\mathcal{D}_N, \mathcal{D}_S).
\]
Set $\xi_{j0} = 0$, $\xi_{j,L_j+1} = 1$ and
\[
  \rho_* = \frac{1}{4} \min_{j \in J_c,\,1 \le \ell \le L_j} \min\{\xi_{j \ell} - \xi_{j,\ell-1}, \xi_{j,\ell+1} - \xi_{j \ell}\},
\]
with $\rho_* = 1$ if $J_c = \varnothing$.

\begin{proof}
The case $B = 1$ follows from $r_{n,1} \ge 1$, $d_1 \le 1$, $|\Delta_{a,1}|\le 2M$, and Theorem~\ref{thm:bgf oracle}. For $B \ge 2$, we first bound $\E d_B^2$, then apply Theorem~\ref{thm:bgf oracle} to obtain the risk bound.

\emph{Step 1: weight extension and empirical bounds.} For $t \ge 1$, let
\[
  \mathfrak{r}_{a,t} = P_{S,a}(v_{a,t} \Phi_t) - Q_S \Phi_t, \qquad \ell_a = \lambda_S/n_{S,a}.
\]
On $\mathcal{C}_n$ the two arm empirical laws are well defined. The first-order variational inequality for \eqref{eq:theory structure refit} is
\begin{equation}
  \label{eq:app refit vi}
  P_{S,a} \left[ \{\Phi_t^\top \mathfrak{r}_{a,t} + \ell_a v_{a,t}\}(u - v_{a,t}) \right] \ge 0
\end{equation}
for every feasible $u$. Holding $\mathfrak{r}_{a,t}$ fixed, this is the variational inequality for the strictly convex empirical problem
\[
  \min_{0 \le u \le C,\ P_{S,a} u=1} P_{S,a} \left\{\frac{\ell_a}{2} u^2 + (\Phi_t^\top \mathfrak{r}_{a,t})u\right\}.
\]
The form of the fitted weights follows directly from the Karush--Kuhn--Tucker conditions. Introduce a multiplier $\alpha_{a,t}$ for the normalization constraint $P_{S,a} u = 1$. Conditional on this multiplier, the KKT conditions minimize pointwise over $z \in [0, C]$ the scalar quadratic
\[
  \frac{\ell_a}{2} z^2 + \{\Phi_t(x)^\top \mathfrak{r}_{a,t} + \alpha_{a,t}\}z.
\]
Its unique minimizer is the Euclidean projection of the unconstrained solution onto $[0, C]$. Writing $\widetilde{c}_{a,t} = -\alpha_{a,t}/\ell_a$ therefore gives
\begin{equation}
  \label{eq:app weight extension}
  \bar{v}_{a,t}(x) = \clip_{[0,C]} \left\{\widetilde{c}_{a,t} - \ell_a^{-1} \Phi_t(x)^\top \mathfrak{r}_{a,t}\right\}.
\end{equation}

Choose $\widetilde{c}_{a,t}$ as the smallest $c$ for which the empirical mean of the clipped expression in \eqref{eq:app weight extension} is at least one. This mean is continuous and nondecreasing in $c$, with limits zero and $C>1$, so the chosen value gives $P_{S,a}\bar{v}_{a,t}=1$. The choice is measurable as a monotone generalized inverse. Strict convexity and \eqref{eq:app refit vi} imply that $\bar{v}_{a,t}$ agrees with $v_{a,t}$ at every arm-$a$ structure observation. Set $\bar{v}_{a,0}=1$ and all extensions to one off $\mathcal{C}_n$. The population comparisons below use these extensions only on $\mathcal{C}_n$.

Because the right-leaf indicator of a stump is one minus its left-leaf indicator, the expression inside the clipping operator in \eqref{eq:app weight extension} is an affine combination of at most $B$ coordinate-stump indicators. Define the underlying affine stump class
\[
  \mathscr{A}_B = \left\{ c + \sum_{b=1}^B \zeta_b \one\{x_{j_b} \le u_b\}: c, \zeta_b \in \mathbb{R}, \ j_b \in [p], \ u_b \in [0, 1] \right\}, \quad \mathscr{V}_B = \clip_{[0,C]}(\mathscr{A}_B).
\]
Equation~\eqref{eq:app weight extension} therefore implies $\bar{v}_{a,t} \in \mathscr{V}_B$ for every $t \le B$. Let $\mathscr{H}$ be the coordinate halfspaces, their complements, and the full cube, and define
\begin{align*}
  \Delta_B & = \max_a \sup_{v \in \mathscr{V}_B,H \in \mathscr{H}} |(P_{S,a} - P_a)(v \one_H)|, \\
  \Delta_Q & = \sup_{H \in \mathscr{H}} |(Q_S - Q)H|, \\
  \Delta_\mu & = \max_a \sup_{H \in \mathscr{H}} |(Q_S - Q)(\widehat{\mu}_a \one_H)|.
\end{align*}

We next bound the size of the entire class of possible fitted weights, including its halfspace restrictions. This step accounts for reuse of the structure sample across all rounds. Fix $B \ge 1$ and $r = A(B + 1)$ points, where the integer $A \ge 2$ depends only on $p$. By adding terms with zero coefficients, we may assume that each sum contains exactly $B$ stump indicators. For a fixed allocation $B_1 + \cdots + B_p = B$, the ordered thresholds on coordinate $j$ have at most $\binom{r + B_j}{B_j}$ patterns. There are at most $(B + 1)^p$ allocations, and
\[
  \prod_{j:B_j > 0} \binom{r + B_j}{B_j} \le \left\{\frac{e(r + B)}{B}\right\}^B \prod_{j:B_j > 0} \left(\frac{B}{B_j}\right)^{B_j} \le \{ep(2A + 1)\}^B.
\]
The last step uses $r + B \le (2A + 1)B$ and $\prod_j (B/B_j)^{B_j} \le p^B$. For each threshold pattern and fixed pseudo-thresholds at the $r$ points, the labels depend on $B + 1$ real coefficients. The hyperplane arrangement bound gives at most
\[
  \sum_{k=0}^{B+1} \binom{r}{k} \le (eA)^{B+1}
\]
label patterns. Clipping does not increase this count since for $0 \le s < C$, $\clip_{[0,C]}(f) > s$ is equivalent to $f > s$. Pseudo-thresholds outside $[0, C)$ give fixed labels.

The class $\{v \one_H:v \in \mathscr{V}_B, H \in \mathscr{H}\}$ needs a further count. There are at most $2p(r + 1) + 1$ membership patterns for $H$ on the same points. Once this pattern is fixed, observations outside $H$ have value zero and those inside $H$ have the same labels as under $v$. Thus both classes have at most
\[
  \{2p(r + 1) + 1\}(B + 1)^p(eA)^{B+1} \{ep(2A + 1)\}^B
\]
pseudo-label patterns. Using $r=A(B+1)$ and $B + 1 \le 2^B$, we bound this by 
\[
  (2p + 1)(A + 1)eA \{2^{p+1} e^2 pA(2A + 1)\}^B.
\]
Choose $A = A_p$ so large that both $(2p + 1)(A + 1)eA < 2^A$ and $2^{p+1} e^2 pA(2A + 1) < 2^A$. The count is then strictly smaller than $2^{A(B+1)}$ for every $B \ge 1$. Write $\operatorname{Pdim}$ for pseudo-dimension. Consequently,
\begin{equation}
  \label{eq:app pdim}
  \operatorname{Pdim}(\mathscr{V}_B) \vee \operatorname{Pdim}\{v \one_H:v \in \mathscr{V}_B, H \in \mathscr{H}\} \le A_p(B + 1).
\end{equation}

Write $N(\eta, \mathscr{G}, L_2(P'))$ for the covering number of $\mathscr{G}$ at radius $\eta$ in $L_2(P')$. For a class $\mathscr{G}$ bounded by $U$ with VC-subgraph dimension $d$, the uniform entropy bound and symmetrization give
\[
  \log N(\epsilon U, \mathscr{G}, L_2(P')) \le K(d + 1) \log(K/\epsilon), \qquad \E Z_q \le KU \sqrt{(d + 1)/q},
\]
where $0 < \epsilon < 1$, $P'$ is any probability law, and $Z_q = \sup_{g \in \mathscr{G}} |(P_q - P)g|$ \citep{vdvaartwellner1996}. Replacing one observation changes $Z_q$ by at most $2U/q$. The Efron--Stein inequality therefore gives $\Var(Z_q) \le 2U^2/q$ \citep{efronstein1981}, and hence
\[
  \E Z_q^2 = (\E Z_q)^2 + \Var(Z_q) \le KU^2(d + 1)/q.
\]
Apply this bound to the deterministic, capped class in \eqref{eq:app pdim}, separately within each arm conditional on the structure labels. For $\Delta_Q$, apply \eqref{eq:bounded halfspace moment} under $Q$ with $h = 1$, without conditioning on the labels. For $\Delta_\mu$, apply it with $h = \widehat{\mu}_a$, conditional only on $\mathcal{D}_N$. Since $n_{S,a} \ge n \pi_*/2$ on $\mathcal{C}_n$,
\begin{equation}
  \label{eq:app empirical second moments}
  \E\{\one_{\mathcal{C}_n} \Delta_B^2\} \le K \frac{B + 1}{n}, \qquad \E\{\one_{\mathcal{C}_n}(\Delta_Q^2 + G^{-2} \Delta_\mu^2)\} \le \frac{K}{n}.
\end{equation}
The first bound is uniform over all $t \le B$ because every fitted extension lies in $\mathscr{V}_B$.

Define
\begin{equation}
  \mathcal{E}_n = \mathcal{A}_N \cap \mathcal{C}_n \cap \{\Delta_B \le 1/2\}.
\end{equation}
Since the full cube belongs to $\mathscr{H}$ and $P_{S,a} \bar{v}_{a,t} = 1$,
\[
  |P_a \bar{v}_{a,t} - 1| \le \Delta_B
\]
uniformly over $a$ and $t \le B$. Hence on $\mathcal{E}_n$,
\begin{equation}
  \label{eq:app population mass}
  \frac{1}{2} \le P_a \bar{v}_{a,t} \le \frac{3}{2}.
\end{equation}
Assumption~\ref{assump:bgf contrast signal}, Hoeffding's bound for $\mathcal{C}_n$, Markov's inequality, and \eqref{eq:app empirical second moments} yield
\begin{equation}
  \label{eq:app mass event prob}
  \Prb(\mathcal{E}_n^c) \le \varepsilon_N + K \frac{B + 1}{n}.
\end{equation}

\emph{Step 2: detecting unresolved jumps.} Fix $t$ with $\lfloor B/2\rfloor \le t < B$ and set 
\[
  r_t = \min(d_t, \rho_*), \qquad \beta = 2 \nu + 2.
\]
If $d_t = 0$, the desired one-step bound below is trivial. Otherwise choose a threshold $\xi = \xi_{j \ell}$ attaining $d_t$. By the definitions of $d_t$ and $\rho_*$, the interval $(\xi - r_t, \xi + r_t)$ contains neither a split threshold selected by iteration $t$ nor another true assignment threshold in coordinate $j$. Since \eqref{eq:app weight extension} is a clipped affine function of the stump indicators in $\Phi_t$, for each fixed $y = x_{-j}$ it is constant in $x_j$ on this interval:
\[
  \bar{v}_{a,t}(u, y) = c_a(y), \qquad0 \le c_a(y) \le C.
\]
Define the centered population discrepancy
\begin{equation}
  D_{a,t,j}(u) = P_a \{\bar{v}_{a,t} \one(X_j \le u)\} - uP_a \bar{v}_{a,t}.
\end{equation}
On $\mathcal{E}_n$, choose the arm $a = a(j)$ supplied by Assumption~\ref{assump:bgf contrast signal}, and write $D(u) = D_{a,t,j}(u)$. Product-uniform $Q$ and $dP_a/dQ = e_a/\pi_a$ give, for almost every $u$ in the interval,
\begin{equation}
  \label{eq:app D derivative}
  D'(u) = \pi_a^{-1} \int e_a(u, y)c_a(y)\,dy - P_a \bar{v}_{a,t}.
\end{equation}
There are two cases. If $\int c_a(y)dy \le \pi_a/4$, then $e_a \le 1$ and \eqref{eq:app population mass} imply $D'(u) \le -1/4$ throughout the interval. Hence $\sup_{|u-\xi| \le r_t} |D(u)| \ge r_t/4$.

Otherwise $\int c_a(y)dy > \pi_a/4$. Write $c_{1,j \ell} = c_{j \ell}$ and $c_{0,j \ell} = -c_{j \ell}$. For $0 < s < r_t$, the points $\xi - r_t + s$ and $\xi + s$ straddle exactly the selected jump, so
\[
  e_a(\xi + s, y) - e_a(\xi - r_t + s, y) = c_{a,j \ell}.
\]
Integrating the difference of \eqref{eq:app D derivative} over $s \in (0, r_t)$ gives
\begin{equation}
  \label{eq:app second difference}
  D(\xi - r_t) - 2D(\xi) + D(\xi + r_t) = \frac{c_{a,j \ell} r_t}{\pi_a} \int c_a(y)dy.
\end{equation}
Its magnitude is at least $\kappa r_t/4$, so at least one of the three values in \eqref{eq:app second difference} has magnitude at least $\kappa r_t/16$. Thus $c_1 = \min\{1/4, \kappa/16\}$ satisfies
\begin{equation}
  \sup_{|u-\xi| \le r_t} |D(u)| \ge c_1 r_t.
\end{equation}
The function $D$ is absolutely continuous, with Lipschitz constant at most $L_D = C/\pi_* + C$. Choose a maximizing point $u_0$ in the closed interval and set $c_2 = \min\{1, c_1/(2L_D)\}$ and $c_3 = c_1/2$. One side of $u_0$ contains a closed interval $I_t$ of length $c_2 r_t$ within $[\xi - r_t, \xi + r_t]$. For $u \in I_t$,
\[
  |D(u)| \ge |D(u_0)| - L_D|u - u_0| \ge c_3 r_t.
\]
This interval lies in $(0, 1)$ because $r_t \le \rho_*$ and the assignment jumps are separated from the endpoints.

Apply Assumption~\ref{assump:bgf contrast signal} to $I_t$. It gives a closed subset $K_t \subset I_t$ of Lebesgue measure at least $\chi c_2 r_t$ on which
\[
  |\widehat{b}_{a(j)j}(u)| \ge c_g(c_2 r_t)^\nu.
\]
Every $u \in K_t$ therefore satisfies $|D(u) \widehat{b}_{a(j)j}(u)| \ge c_3 c_g c_2^\nu r_t^{\nu+1}$. Consequently, if
\[
  \Gamma_t^P(j, u) = \sum_{a=0}^1 \{D_{a,t,j}(u) \widehat{b}_{aj}(u)\}^2,
\]
then
\begin{equation}
  \sup_{0 < u < 1} \Gamma_t^P(j, u) \ge c r_t^\beta.
\end{equation}

\emph{Step 3: threshold recovery.} For $0 < q < 1$, define
\[
  \mathcal{P}(q, d, z) = \frac{dz}{q(1 - q)}.
\]
This parametrization is exactly the product entering one arm of the split score. Indeed, for an empirical halfspace $H = H_{j,u}$ with $q = Q_S H \in (0, 1)$, let
\[
  A = P_{S,a}(v_{a,t} \one_H), \qquad d = A - q, \qquad z = Q_S(\widehat{\mu}_a \one_H) - qQ_S \widehat{\mu}_a.
\]
Since $P_{S,a} v_{a,t} = 1$, we have $d = (1 - q)A - q(1 - A)$.

On $\mathcal{C}_n$, the number of arm-$a$ observations in $H$ is at most $nq$, so
\[
  P_{S,a} H \le \frac{nq}{n_{S,a}} \le \frac{2q}{\pi_*}, \qquad P_{S,a} H^c \le \frac{2(1 - q)}{\pi_*}.
\]
Using $0 \le v_{a,t} \le C$ gives
\[
  A \le \frac{2Cq}{\pi_*}, \qquad 1 - A = P_{S,a}(v_{a,t} \one_{H^c}) \le \frac{2C(1 - q)}{\pi_*},
\]
Both $(1 - q)A$ and $q(1 - A)$ are nonnegative and at most $(2C/\pi_*)q(1 - q)$. Their difference therefore satisfies
\begin{equation}
  |d| \le Wq(1 - q), \qquad W: = \frac{2C}{\pi_*}.
\end{equation}
Also
\[
  z = (1 - q) Q_S(\widehat{\mu}_a \one_H) - q Q_S(\widehat{\mu}_a \one_{H^c}),
\]
so $|z| \le 2G q(1 - q)$.

The population triple satisfies the same bounds. Write $A^P = P_a(\bar{v}_{a,t} \one_H)$ and $M^P = P_a \bar{v}_{a,t}$. Since
\[
  d^P = D_{a,t,j}(u) = A^P - u M^P= (1 - u)A^P - u(M^P - A^P)
\]
and $dP_a/dQ = e_a/\pi_a \le 1/\pi_*$,
\[
  A^P \le \frac{Cu}{\pi_*}, \qquad M^P - A^P \le \frac{C(1 - u)}{\pi_*},
\]
so $|d^P| \le (C/\pi_*)u(1 - u) \le Wu(1 - u)$. The population centered $\widehat{\mu}_a$ moment
\[
  z^P = Q(\widehat{\mu}_a \one_H) - u Q \widehat{\mu}_a
\]
satisfies $|z^P| \le 2Gu(1 - u)$ by the same decomposition into $H$ and $H^c$.

On the domain
\[
  \mathscr{D} = \{(q, d, z): 0<q<1, |d| \le Wq(1 - q), \ |z| \le 2G q(1 - q)\},
\]
direct differentiation gives
\[
  |\partial_d \mathcal{P}| \le 2G, \qquad |\partial_z \mathcal{P}| \le W, \qquad |\partial_q \mathcal{P}| \le 2G W.
\]
The domain $\mathscr{D}$ is convex because $q(1-q)$ is concave, each constraint $d \le W q(1-q)$, $-d \le W q(1-q)$ and the two analogous constraints for $z$ define a convex set. Hence the derivative bounds hold along the segment joining any empirical and population triples in $\mathscr{D}$.

For the same halfspace $H = H_{j,u}$, the definitions of $\Delta_B, \Delta_Q, \Delta_\mu$ give
\begin{align}
  |Q_S H - u| & \le \Delta_Q, \label{eq:app q comparison} \\
  |I_{a,t}(j, u) - D_{a,t,j}(u)| & \le 2 \Delta_B + \Delta_Q, \label{eq:app d comparison} \\
  \left| Q_S(\widehat{\mu}_a \one_H) - Q_S H\,Q_S \widehat{\mu}_a - \{Q(\widehat{\mu}_a \one_H) - uQ \widehat{\mu}_a\} \right| & \le 2 \Delta_\mu + G \Delta_Q.
\end{align}

For \eqref{eq:app d comparison}, use that $v_{a,t}$ and $\bar{v}_{a,t}$ agree on the structure observations and $|P_a \bar{v}_{a,t} - 1| \le \Delta_B$. The mean-value theorem applied arm by arm to $\mathcal{P}$, followed by the reverse triangle inequality for the Euclidean norm over the two arms, yields
\begin{equation}
  \label{eq:app score comparison}
  |\sqrt{\Gamma_{t+1}^{\mathrm{emp}}(j, u)} - \sqrt{\Gamma_t^P(j, u)}| \le K\{G(\Delta_B + \Delta_Q) + \Delta_\mu\},
\end{equation}
whenever the empirical split induced by $u$ has two nonempty children. Here $\Gamma_{t+1}^{\mathrm{emp}}(j, u)$ denotes the empirical score computed from that sample partition. If a cutoff in $\mathcal{U}_j$ induces the same partition, it has exactly this score.

We now connect the population cut found in Step 2 to an admissible empirical cutoff. Define the maximal empirical score over all coordinates and admissible cutoffs by
\[
  \Gamma_{t+1}^* = \max_{j \in [p]} \max_{u \in \mathcal{U}_j} \Gamma_{t+1}(j, u).
\]
For each realization choose any $u_t^\star \in K_t$ from Step 2. The following comparisons hold uniformly over cutoffs and hence do not require a measurable selection of this point. In particular,
\begin{equation}
  \label{eq:app population discovery point}
  \Gamma_t^P(j, u_t^\star) \ge c r_t^\beta.
\end{equation}
Set $q_t = Q_S H_{j,u_t^\star}$ and $\ell_t = nq_t$, which is an integer because $Q_S$ is empirical. There are two cases.

If $m \le \ell_t \le n - m$, continuity of $Q$ implies that the structure covariates have no ties almost surely. Hence there is a cutoff $\widetilde{u}_t \in \mathcal{U}_j$ between the $\ell_t$th and $(\ell_t + 1)$st ordered observations that induces exactly the same sample partition as $u_t^\star$. Thus $\Gamma_{t+1}(j, \widetilde{u}_t) = \Gamma_{t+1}^{\mathrm{emp}}(j, u_t^\star)$. By \eqref{eq:app score comparison},
\[
  \sqrt{\Gamma_{t+1}(j, \widetilde{u}_t)} \ge \left[ \sqrt{\Gamma_t^P(j, u_t^\star)} - K\{G(\Delta_B + \Delta_Q) + \Delta_\mu\} \right]_ + .
\]
Using $(x - y)_ + ^2 \ge x^2/2 - y^2$ and \eqref{eq:app population discovery point} gives
\begin{equation}
  \label{eq:app admissible discovery}
  \Gamma_{t+1}^* \ge c r_t^\beta - K\{G^2(\Delta_B^2 + \Delta_Q^2) + \Delta_\mu^2\}.
\end{equation}

If instead $\ell_t < m$ or $n - \ell_t < m$, then $\min(q_t, 1 - q_t) < m/n$. By \eqref{eq:app q comparison},
\[
  \min(u_t^\star, 1 - u_t^\star) \le \frac{m}{n} + \Delta_Q.
\]
The population-domain bounds above and $|\widehat{b}_{aj}(u)| \le 2G$ imply
\[
  \sqrt{\Gamma_t^P(j, u_t^\star)} \le KG \min(u_t^\star, 1 - u_t^\star) \le KG \left(\frac{m}{n} + \Delta_Q\right).
\]
Combining this with \eqref{eq:app population discovery point} gives
\[
  cr_t^\beta \le KG^2 \left\{(m/n)^2 + \Delta_Q^2\right\}.
\]
Since $\Gamma_{t+1}^* \ge 0$, after reducing $c$ and enlarging $K$ this implies the same lower bound as \eqref{eq:app admissible discovery}, with the additional $G^2(m/n)^2$ error.

Thus, in both cases, on $\mathcal{E}_n$,
\begin{equation}
  \label{eq:app empirical discovery}
  \Gamma_{t+1}^* \ge cr_t^\beta - \mathfrak{e}_n,
\end{equation}
where
\begin{equation}
  \mathfrak{e}_n = K \left\{ G^2(\Delta_B^2 + \Delta_Q^2) + \Delta_\mu^2 + G^2(m/n)^2 \right\}.
\end{equation}
By \eqref{eq:app empirical second moments},
\begin{equation}
  \label{eq:app mean comparison error}
  \E\{\one_{\mathcal{C}_n} \mathfrak{e}_n\} \le KG^2 \left\{\frac{B}{n} + \left(\frac{m}{n}\right)^2\right\}.
\end{equation}

Before the coordinate subset at iteration $t + 1$ is drawn, $v_t$, $\Gamma_{t+1}^*$, $r_t$, $\mathfrak{e}_n$, and $\mathcal{E}_n$ are $\mathcal{F}_t$-measurable. Choose the smallest coordinate among those attaining $\Gamma_{t+1}^*$. The fresh uniform size-$m_{\mathrm{try}}$ set $\mathcal{M}_{t+1}$ contains this coordinate with conditional probability exactly $m_{\mathrm{try}}/p$. On that event the maximization performed by BGF over the offered coordinates and all their admissible cutoffs has value at least $\Gamma_{t+1}^*$. Since scores are nonnegative on the complementary event,
\begin{equation}
  \E(S_{t+1} \mid \mathcal{F}_t) \ge \frac{m_{\mathrm{try}}}{p} \Gamma_{t+1}^*.
\end{equation}
Multiplying this conditional inequality by the $\mathcal{F}_t$-measurable indicator $\one_{\mathcal{E}_n}$, using \eqref{eq:app empirical discovery}, and taking expectations gives
\begin{equation}
  \label{eq:app one step resolution}
  c \E(\one_{\mathcal{E}_n} r_t^\beta) \le \frac{p}{m_{\mathrm{try}}} \E S_{t+1} + K \E(\one_{\mathcal{C}_n} \mathfrak{e}_n).
\end{equation}
The comparison error is not multiplied by $p/m_{\mathrm{try}}$: it enters \eqref{eq:app empirical discovery} before the fresh-coordinate probability is inverted.

Let $t_0 = \lfloor B/2\rfloor$ and $L = B - t_0$. Average \eqref{eq:app one step resolution} over $t_0 \le t < B$. Proposition~\ref{prop:bgf imbalance} with $T = B$ and \eqref{eq:app mean comparison error} give
\begin{equation}
  \frac{1}{L} \sum_{t=t_0}^{B-1} \E(\one_{\mathcal{E}_n} r_t^\beta) \le Kr_{n,B}.
\end{equation}
Since $Z_{j,t} \subseteq Z_{j,B}$ for every $j$ and $t \le B$, $d_B \le d_t$ and therefore $r_B: = \min(d_B, \rho_*) \le r_t$. Hence, 
\begin{equation}
  \label{eq:app terminal beta moment}
  \E(\one_{\mathcal{E}_n} r_B^\beta) \le Kr_{n,B}.
\end{equation}
Since $0 < \rho_* \le 1$ and $d_B \le 1$, the truncation obeys $d_B \le r_B/\rho_*$. For $\beta = 2 \nu + 2 \ge 2$, Lyapunov's inequality and \eqref{eq:app terminal beta moment} give
\[
  \E(\one_{\mathcal{E}_n} d_B^2) \le \rho_*^{-2} \{\E(\one_{\mathcal{E}_n} r_B^\beta)\}^{2/\beta} \le K \rho_*^{-2} r_{n,B}^{1/(\nu+1)}.
\]
The separation $\rho_*$ is fixed under Assumption~\ref{assump:bgf jumps}. Absorbing its reciprocal in $K$ is therefore valid for this fixed assignment law. A uniform rate over varying assignment laws would require a common positive separation bound. On $\mathcal{E}_n^c$, $d_B^2 \le 1$, so \eqref{eq:app mass event prob} gives a contribution at most $\varepsilon_N + K(B + 1)/n$. Since $B \ge 2$, $m_{\mathrm{try}} \le p$, and $\lambda_S > 0$, the definition of $r_{n,B}$ implies $(B + 1)/n \le Kr_{n,B}$. If $r_{n,B} \le 1$, then $r_{n,B} \le r_{n,B}^{1/(\nu+1)}$. If $r_{n,B} > 1$, the desired bound follows from $d_B^2 \le 1$. Combining the good and exceptional events therefore yields
\[
  \E d_B^2 \le K\{\varepsilon_N + r_{n,B}^{1/(\nu+1)}\}.
\]
This proves
\begin{equation}
  \label{eq:app threshold recovery}
  \E d_B^2 \le K \rho_{n,B}.
\end{equation}

\emph{Step 4: ATE risk bound.} Condition on the realized stump forest and let $\mathscr{F} = \sigma(\Phi_B(X))$. For each coordinate $j$, let $\mathscr{F}_j$ be generated by the split thresholds of the forest in that coordinate and let $\Pi_j$ denote conditional expectation under $Q$ given $\mathscr{F}_j$. Note that the forest cells are rectangles and $Q$ is a product measure, so the coordinates remain independent conditionally on $\mathscr{F}$.

For $j \in J_e$, with $c_{1,j \ell} = c_{j \ell}$ and $c_{0,j \ell} = -c_{j \ell}$, define
\[
  e_{a,j}(x_j) = \sum_{\ell=1}^{L_j} c_{a,j \ell} \one(x_j > \xi_{j \ell}),
\]
and set $e_{a,j} \equiv 0$ for $j \notin J_e$. The intercepts of $e_1$ and $e_0$ are $\alpha_0$ and $1 - \alpha_0$, respectively, and disappear upon subtracting the cellwise conditional means. Together with additivity of $\mu_a$, this gives
\[
  e_a - e_{a,B}^Q = \sum_{j=1}^p (e_{a,j} - \Pi_j e_{a,j}), \qquad \mu_a - \mu_{a,B}^Q = \sum_{j=1}^p (f_{aj} - \Pi_j f_{aj}).
\]

For $j \ne k$, conditional independence of the coordinate cells and the conditional-centering identities $\E_Q(e_{a,j} - \Pi_j e_{a,j} \mid \mathscr{F}_j) = 0$ and $\E_Q(f_{ak} - \Pi_k f_{ak} \mid \mathscr{F}_k) = 0$ imply
\[
  \E_Q \{(e_{a,j} - \Pi_j e_{a,j})(f_{ak} - \Pi_k f_{ak}) \mid \mathscr{F}\} = 0.
\]
Consequently the conditional covariance identity \eqref{eq:app covariance identity} reduces to
\[
  e_{a,B}^Q \Delta_{a,B} = \sum_j \E_Q \{(e_{a,j} - \Pi_j e_{a,j})(f_{aj} - \Pi_j f_{aj}) \mid \mathscr{F}\}.
\]
Because $\Delta_{a,B}$ is $\mathscr{F}$-measurable,
\[
  P_a|\Delta_{a,B}| = \frac{1}{\pi_a} Q(e_a|\Delta_{a,B}|) = \frac{1}{\pi_a} Q(e_{a,B}^Q|\Delta_{a,B}|).
\]
Take absolute values in the preceding conditional-covariance display, integrate under $Q$, and use conditional Jensen. Coordinates outside $J_c$ have either no assignment component or no outcome component, so their diagonal term is zero. Hence
\begin{equation}
  P_a|\Delta_{a,B}| \le \frac{1}{\pi_a} \sum_{j \in J_c} Q \left|(e_{a,j} - \Pi_j e_{a,j})(f_{aj} - \Pi_j f_{aj})\right|.
\end{equation}
Because the $f_{aj}$ are centered and $|\mu_a| \le M$,
\[
  f_{aj}(x_j) = \E_Q \{\mu_a(X) \mid X_j = x_j\} - Q \mu_a,
\]
so $|f_{aj}| \le 2M$ and $|f_{aj} - \Pi_j f_{aj}| \le 4M$. Thus
\begin{equation}
  P_a|\Delta_{a,B}| \le \frac{4M}{\pi_a} \sum_{j \in J_c} Q|e_{a,j} - \Pi_j e_{a,j}|.
\end{equation}
For each true threshold, choose a nearest point $z_{j \ell}$ in $Z_{j,B}$ and set
\[
  \widetilde{e}_{a,j}(x_j) = \sum_{\ell=1}^{L_j} c_{a,j \ell} \one(x_j > z_{j \ell}).
\]
This function is $\mathscr{F}_j$-measurable. Since $X_j$ is uniform, we have 
\[
  Q|e_{a,j} - \widetilde{e}_{a,j}| \le \sum_{\ell=1}^{L_j} |c_{j \ell}|\,|\xi_{j \ell} - z_{j \ell}|.
\]

Furthermore, $\widetilde{e}_{a,j} - \Pi_j e_{a,j} = \Pi_j(\widetilde{e}_{a,j} - e_{a,j})$, so conditional Jensen gives
\[
  Q |e_{a,j} - \Pi_j e_{a,j}| \le 2 Q |e_{a,j} - \widetilde{e}_{a,j}|.
\]
Together with 
\[
  V_e = \sum_{j \in J_c} \sum_{\ell=1}^{L_j} |c_{j \ell}|,
\]
we obtain
\begin{equation}
  \label{eq:app delta distance}
  P_a|\Delta_{a,B}| \le KMV_e d_B.
\end{equation}
The quantity $V_e$ is fixed under Assumption~\ref{assump:bgf jumps}. Apply Theorem~\ref{thm:bgf oracle}, conditionally on $\mathcal{F}_B$, and use \eqref{eq:app delta distance}. Then we have 
\[
  \E\{(\widehat{\theta} - \theta)^2 \mid \mathcal{F}_B\} \le K \left\{ \frac{M^2 + \sigma_Y^2}{n} + \frac{BR^2}{n} + M^2 d_B^2 \right\}.
\]
Taking expectations and applying \eqref{eq:app threshold recovery} gives \eqref{eq:bgf consistency risk}, completing the proof.
\end{proof}

\section{Identities for the balancing and splitting criteria}
\label{app:score identities}

The identities below explain Sections~\ref{sec:forest balancing}--\ref{sec:bgf}. They hold for the stated fixed weights or realized partitions.

\subsection{Conditional risk and the fixed-feature problem}

Condition on an independently constructed representation and on the analysis covariates and treatment labels. Suppose both arms are nonempty, the weights depend only on these conditioned quantities, and $\sigma_a^2(x) = \Var(Y \mid X = x, A = a) < \infty$. Define $\varepsilon_i = Y_i - \mu_{A_i}(X_i)$. Conditional independence of the observations gives
\begin{equation}
  \begin{aligned}
    \E \left[\left\{\sum_{i \in \mathcal{I}_a} w_i Y_i - Q_n \mu_a\right\}^2 \Bigm|X_{1:n}, A_{1:n}, \Phi\right] & = \{(\widehat{P}_{a,w} - Q_n) \mu_a\}^2 + \sum_{i \in \mathcal{I}_a} w_i^2 \sigma_a^2(X_i).
  \end{aligned}
  \label{eq:conditional risk exact}
\end{equation}
Indeed, the expression inside braces equals $(\widehat{P}_{a,w} - Q_n) \mu_a + \sum_i w_i \varepsilon_i$. The last term has conditional mean zero and conditional variance $\sum_i w_i^2 \sigma_a^2(X_i)$. The target in \eqref{eq:conditional risk exact} is the sample covariate mean $Q_n \mu_a$. For the population target, replace the bias by $(\widehat{P}_{a,w} - Q_n) \mu_a + (Q_n - Q) \mu_a$. The latter sampling term is also accounted for by Theorem~\ref{thm:bgf oracle}.

For $\mu_a = c + \beta^\top \Phi + u_a$, normalization gives the exact decomposition
\[
  (\widehat{P}_{a,w} - Q_n) \mu_a = \beta^\top r_a(w; \Phi) + (\widehat{P}_{a,w} - Q_n)u_a.
\]
Cauchy--Schwarz, attained by $\beta$ proportional to $r_a$ when $r_a \ne 0$, proves \eqref{eq:function class imbalance}. If $u_a = 0$, $\|\beta\|_2 \le R$ and $\sigma_a^2 \le \sigma^2$, the conditional risk is bounded by $R^2\|r_a\|_2^2 + \sigma^2\|w\|_2^2$. This is the bias--variance interpretation of balancing weights \citep{kallus2020,brunssmith2022}. The penalty parameter $\lambda$ tunes this trade-off.

For $g = c + \beta^\top \Phi$, normalization gives $\widehat{P}_{a,w} g - Q_n g = \beta^\top r_a(w; \Phi)$. Adding and subtracting $\widehat{P}_{a,w} g$ proves \eqref{eq:calibration regression}. The first two terms in that identity, $Q_n g + \widehat{P}_{a,w}(Y - g)$, change by $-\varepsilon(\widehat{P}_{a,w} - Q_n)h$ when $g$ is replaced by $g + \varepsilon h$.

Let $d$ denote the dimension of $\Phi$. The same refit has the exact regularized saddle representation
\[
  \min_{w \in \mathcal{W}_a} \max_{\beta \in \mathbb{R}^d} \left\{2(\widehat{P}_{a,w} - Q_n)(\beta^\top \Phi) - \|\beta\|_2^2 + \lambda\|w\|_2^2\right\}.
\]
For each fixed $w$, completing the square shows that the inner maximum is attained at $\beta = r_a(w; \Phi)$, with value $\|r_a(w; \Phi)\|_2^2 + \lambda\|w\|_2^2$. 

For completeness, the kernel implementation is
\begin{equation}
  \|r_a(w; \Phi)\|_2^2 = w^\top \Kbm_{aa} w - 2w^\top \bar{k}_{a,Q} + c_Q,
  \label{eq:kernel mmd}
\end{equation}
where $k(x, x') = \Phi(x)^\top \Phi(x')$, $\Kbm_{aa} = \allowbreak [k(X_i, X_j)]_{i,j \in \mathcal{I}_a}$, $(\bar{k}_{a,Q})_i = n^{-1} \sum_j k(X_i, X_j)$, and $c_Q = n^{-2} \sum_{i,j} k(X_i, X_j)$. Expanding the squared norm proves \eqref{eq:kernel mmd}. Applying the same expansion to concatenated leaf indicators proves \eqref{eq:leaf balance}. Finally, $0 \le w_i \le C/n_a$ and $\sum_i w_i = 1$ imply $\sum_i w_i^2 \le (C/n_a) \sum_i w_i = C/n_a$, so the effective sample size $1/\sum_i w_i^2$ is at least $n_a/C$.

\subsection{Projection refinement, prediction gain, and weighting bias}

Fix a split with $Q(L), Q(R) > 0$ and $\mu \in L_2(Q)$. Since $\mu_N = q \mu_L + (1 - q) \mu_R$, the refinement in \eqref{eq:centered refinement} satisfies $Q h_{L;N} = 0$ and
\begin{equation}
  Q\{\one_N(\mu - \mu_N)^2\} - Q\{\one_L(\mu - \mu_L)^2 + \one_R(\mu - \mu_R)^2\} = Qh_{L;N}^2 = Q(N)q(1 - q)(\mu_L - \mu_R)^2.
  \label{eq:cart gain exact}
\end{equation}

To verify the first equality, expand the two squares on each child. The mixed term vanishes because $Q\{\one_D(\mu - \mu_D)\} = 0$ for $D = L, R$. The remaining term is $Q(L)(\mu_L - \mu_N)^2 + Q(R)(\mu_R - \mu_N)^2$, which simplifies to the displayed value. Applying $P_w - Q$ to $h_{L;N}$ gives $\delta_w(L; N)(\mu_L - \mu_R)$. Squaring and substituting \eqref{eq:cart gain exact} proves \eqref{eq:cart factorization}. Under Assumption~\ref{assump:identification} and finite second moments, the same prediction gain is obtained for $Y(a)$. The conditional variance term is the same under the parent and child fits and cancels when their risks are subtracted.

For the normalized contrast in Section~\ref{sec:population score}, $Q \psi_{L;N} = 0$ and $Q \psi_{L;N}^2 = 1$ by direct integration. Furthermore,
\[
  Q(\mu \psi_{L;N}) = \{Q(N)q(1 - q)\}^{1/2}(\mu_L - \mu_R), \qquad Q(\rho_w \psi_{L;N}) = \frac{\delta_w(L; N)}{\{Q(N)q(1 - q)\}^{1/2}}.
\]
Multiplying proves \eqref{eq:directional score}. In the treatment-arm problem, $dP_w/dQ = we_a/\pi_a$, which gives the stated density error.

For a finite realized tree $T$, let $\mathcal{N}(T)$ be its internal nodes and $\Pi_T \mu = \sum_{\ell \in \mathcal{L}(T)} \mu_\ell \one_\ell$. Assume all leaves have positive $Q$-mass and the root is $\mathcal{X}$. The normalized contrasts for different nodes are orthogonal. Contrasts at disjoint nodes have disjoint support. When the nodes are nested, the ancestor contrast is constant on the descendant and the descendant has zero mean. Telescoping the changes induced by successive splits gives 
\[
  \Pi_T \mu = Q \mu + \sum_{N \in \mathcal{N}(T)} h_{L_N;N}, \qquad (P_w - Q) \Pi_T \mu = \sum_{N \in \mathcal{N}(T)} \delta_w(L_N; N)(\mu_{L_N} - \mu_{R_N}).
\]
This proves \eqref{eq:tree bias expansion}. Total squared bias contains cross-products between signed node contributions. The score is the square of the bias contribution from a single split. The analysis of repeated refitting in Section~\ref{sec:theory feedback} is therefore needed in addition to the projection identity.

For the ATE let $b_a = (P_{a,w_a} - Q)h_{a,L;N}$. The inequality $(b_1 - b_0)^2 \le 2(b_1^2 + b_0^2)$ follows from $2|b_1 b_0| \le b_1^2 + b_0^2$. This explains the sum of squares in \eqref{eq:ate score}. The empirical one-arm score, using the same definitions as Section~\ref{sec:bgf ate}, is 
\begin{equation}
  \widehat{G}_{a,w}(L; N) = \{\omega_a(L) - \widehat{q}\,\omega_a(N)\}^2(\bar{\mu}_{a,L} - \bar{\mu}_{a,R})^2.
\end{equation}

\subsection{Outcome contrasts under the target law}

For a measurable region $D$ with $Q(D) > 0$, change of measure gives
\[
  \begin{aligned}
    \E(Y \mid A = a, X \in D) & = \frac{Q(e_a \mu_a \mid D)}{Q(e_a \mid D)}, \\
    \E(Y \mid A = a, X \in D) - Q(\mu_a \mid D) & = \frac{\Cov_Q(e_a, \mu_a \mid D)}{Q(e_a \mid D)}.
  \end{aligned}
\]
The denominator is positive under overlap. Subtraction of $Q(\mu_a \mid D)$ proves \eqref{eq:node selection covariance}. This identity explains why BGF averages predictions over the target covariates instead of using raw within-arm node means. Replacing $\mu_a$ by $\widehat{\mu}_a$ introduces outcome regression error.

For fixed candidate splits, replacing $\widehat{\mu}_a$ by $c \widehat{\mu}_a + d_a$, with the same $c \ne 0$ for both arms, multiplies every score by $c^2$. With fixed randomization, tie-breaking, growth parameters and penalties, the entire sequence of partitions and refitted weights is therefore unchanged, by induction over trees and their nodes.

A special case of the pure-interaction design in Appendix~\ref{app:pure interaction stress test} illustrates the limitation. Let $X_1, X_2$ be independent standard normal variables, $U = \operatorname{sign}(X_1 X_2)$, $e_1 = 1/2 + U/4$, and $\mu_a = U + a \theta$. Then $Q(U \mid X_j) = 0$ for $j = 1, 2$, so every coordinate root split has zero outcome contrast even when $\widehat{\mu}_a = \mu_a$ for both arms. In this example, $\pi_1 = 1/2$, $P_1 U = 1/2$, and $P_0 U = -1/2$, so uniform weights give $\theta(w) - \theta = 1$. A split on the sign of one coordinate can expose a contrast in the other coordinate, which deeper trees may use. Small root scores alone therefore cannot certify absence of confounding.

\subsection{Identities for calibrated weight feedback}
\label{app:feedback identities}

All empirical statements in this subsection condition on a realized structure sample with $n_a > 0$ and on the successive tree partitions. The partitions may depend on all preceding weight refits. The feasible set $\calW_a$ and $\lambda_S > 0$ remain fixed. This convention matches the actual construction algorithm.

\lead{The discrepancy function and the weights.}
Write $r = r_a(w_a^{(t)}; \Phi_t)$ and $f(x) = \Phi_t(x)^\top r$. The gradient of the objective in \eqref{eq:ate refit} at observation $i$ is $2f(X_i) + 2 \lambda_S w_{a,i}^{(t)}$. If $C > 1$, uniform weights are strictly inside the box relative to the normalization hyperplane. A multiplier $\eta$ for normalization and nonnegative multipliers for the two box constraints therefore satisfy the KKT conditions. Minimizing the coordinate Lagrangian over $z \in [0, C/n_a]$ gives
\[
  \argmin_{0 \le z \le C/n_a} \{\lambda_S z^2 + (2f(X_i) + \eta)z\} = \clip_{[0, C/n_a]} \{ - \eta/(2 \lambda_S) - f(X_i)/\lambda_S \}.
\]
Taking $c_{a,t} = -\eta/(2 \lambda_S)$ proves \eqref{eq:feedback witness}. A normalization multiplier exists because the sum of these clipped values is a continuous nondecreasing function of $c$, with range $[0, C]$. Moreover,
\[
  f(x) = (\widehat{P}_{a,w_a^{(t)}} - Q_n)k_t(x, \cdot) = \frac{1}{t} \sum_{b=1}^t \sum_{\ell \in \calL(T_b)} \{\widehat{P}_{a,w_a^{(t)}}(\ell) - Q_n(\ell)\}\one_\ell(x).
\]
Thus every tree built so far contributes to the current discrepancy function. This is the probability-weight version of the mean-one KKT formula \eqref{eq:app weight extension} used in the threshold-recovery proof. The case $C = 1$ has the unique feasible vector $w_i = 1/n_a$.

\lead{A single stump.}
For $H, H^c$, define $\widehat{p}_a = n_a^{-1} \sum_{i \in \mathcal{I}_a} \one_H(X_i) \in (0, 1)$ and $q = Q_n H$. At fixed arm mass $u$ on $H$, Jensen's inequality shows that the smallest squared weight norm is attained by equal weights within each child:
\[
  \min \sum_{i \in \mathcal{I}_a} w_i^2 = \frac{u^2}{n_a \widehat{p}_a} + \frac{(1 - u)^2}{n_a(1 - \widehat{p}_a)}.
\]
The box constraints are feasible for that mass exactly when
\[
  u \in I_a: = \big[\max\{0, 1 - C(1 - \widehat{p}_a)\}, \, \min\{1, C \widehat{p}_a\}\big].
\]
The two leaf discrepancies are $u - q$ and $-(u - q)$. Therefore the first refit reduces exactly to
\[
  \min_{u \in I_a} \left[ 2(u - q)^2 + \frac{\lambda_S}{n_a} \left\{\frac{u^2}{\widehat{p}_a} + \frac{(1 - u)^2}{1 - \widehat{p}_a}\right\}\right].
\]
The derivative of this quadratic is $4(u - q) + 2 \lambda_S(u - \widehat{p}_a)/\{n_a \widehat{p}_a(1 - \widehat{p}_a)\}$. Its unconstrained minimizer and constrained minimizer are
\[
  u_a^{\mathrm{free}} = \frac{2n_a \widehat{p}_a(1 - \widehat{p}_a)q + \lambda_S \widehat{p}_a}{2n_a \widehat{p}_a(1 - \widehat{p}_a) + \lambda_S}, \qquad u_a = \clip_{I_a}(u_a^{\mathrm{free}}).
\]
When the cap is inactive, subtraction of $q$ proves \eqref{eq:stump feedback shrinkage}. Keeping the fitted outcome regression and the candidate split fixed, the original score is $(\widehat{p}_a - q)^2(\bar{\mu}_{a,H} - \bar{\mu}_{a,H^c})^2$. Its value after this first refit is exactly that score multiplied by $\{\lambda_S/[2n_a \widehat{p}_a(1 - \widehat{p}_a) + \lambda_S]\}^2$. This calculation concerns the first full-sample stump refit. Other splits and later refits can change individual leaf discrepancies. The sequence bound below handles those changes.

\lead{The Riesz directional derivative.}
For $w_a^\star \in L_2(P_a)$, $Qv = P_a(w_a^\star v)$ for $v \in L_2(P_a)$. Consequently,
\[
  \mathcal{R}_a(v) - \mathcal{R}_a(w_a^\star) = P_a(v - w_a^\star)^2.
\]
For $v, h \in L_2(P_a)$, expanding the quadratic gives
\[
  \mathcal{R}_a(v + \varepsilon h) - \mathcal{R}_a(v) = 2 \varepsilon\{P_a(vh) - Qh\} + \varepsilon^2 P_a h^2.
\]
At $v = w_a$, the derivative is $2(P_{a,w_a} - Q)h$. Substitution of $h = h_{a,L;N}$ proves \eqref{eq:riesz directional score}. Substitution of $h = Q(\mu_a \psi_{L;N}) \psi_{L;N}$ gives its outcome-coefficient interpretation. The same derivative identity holds with $\mu_a$ replaced by the fitted outcome regression $\widehat{\mu}_a$, wherever the displayed moments exist. The derivative is taken in the unconstrained space $L_2(P_a)$, and the weights are updated by solving the constrained problem \eqref{eq:ate refit}.

For a full structure-sample split, the empirical identity can also be written directly. Choose any function $v$ agreeing with $n_a w_{a,i}$ on arm-$a$ observations, and define $\mathcal{R}_{S,a}(v) = P_{S,a} v^2 - 2Q_S v$. Then
\[
  \tfrac{1}{2} D \mathcal{R}_{S,a}(v)[h] = P_{S,a}(vh) - Q_S h = (\widehat{P}_{a,w_a} - Q_S)h.
\]
The right side does not depend on how $v$ was extended outside that arm. Taking $h = (\bar{\mu}_{a,L} - \bar{\mu}_{a,R})(\one_L - \widehat{q} \one_N)$ recovers the one-arm empirical score after squaring.

\lead{The cumulative refit inequality.}
Suppress the arm index and write $w^{(t)}$ for its fitted vector. For a realized sequence of trees define
\[
  F_t(v) = \sum_{b=1}^t \ell_b(v) + t \lambda_S\|v\|_2^2, \qquad F_0 = 0,
\]
where $\ell_b(v) = \|r_a(v; \phi_{T_b})\|_2^2$. The averaged feature normalization gives $F_t(v) = t\{\|r_a(v; \Phi_t)\|_2^2 + \lambda_S\|v\|_2^2\}$. Thus $w^{(t)}$ minimizes $F_t$ on the same convex feasible set at every iteration.

For $t \ge 2$, let $d_t = w^{(t)} - w^{(t-1)}$. A quadratic expansion gives
\[
  \begin{aligned}
    F_{t-1}(w^{(t)}) - F_{t-1}(w^{(t-1)}) = {} & \nabla F_{t-1}(w^{(t-1)})^\top d_t + (t - 1) \lambda_S\|d_t\|_2^2\\
    & +\sum_{b=1}^{t-1} \left\|\sum_{i \in \mathcal{I}_a} d_{t,i} \phi_{T_b}(X_i)\right\|_2^2.
  \end{aligned}
\]
The first term is nonnegative by constrained optimality, since the segment from $w^{(t-1)}$ to $w^{(t)}$ is feasible. Hence
\[
  F_t(w^{(t)}) - F_{t-1}(w^{(t-1)}) \ge (t - 1) \lambda_S\|d_t\|_2^2 + \ell_t(w^{(t)}) + \lambda_S\|w^{(t)}\|_2^2.
\]
For $t = 1$, the same inequality holds as equality with the movement term omitted. More explicitly, telescoping gives
\[
  \begin{aligned}
    F_B(w^{(B)}) = {} & \sum_{t=1}^B \{\ell_t(w^{(t)}) + \lambda_S\|w^{(t)}\|_2^2\}\\
    & +\sum_{t=2}^B \{F_{t-1}(w^{(t)}) - F_{t-1}(w^{(t-1)})\}.
  \end{aligned}
\]
The second sum is bounded below by the weighted squared movements. Optimality gives $F_B(w^{(B)}) \le F_B(u)$ for any $u \in \calW_a$, proving \eqref{eq:feedback comparator budget}. The bound holds for every realized tree sequence and controls losses evaluated \emph{after} each refit. The argument is a quadratic, constrained version of the leader comparison used in online learning \citep{shalev2012online}. 

To connect this bound to the actual pre-refit score, consider full-sample stumps and $|\widehat{\mu}_a| \le G$. For the selected stump $T_{t+1}$,
\[
  r_a(w_a^{(t)}; \phi_{T_{t+1}}) = r_a(w_a^{(t+1)}; \phi_{T_{t+1}}) - \sum_{i \in \mathcal{I}_a} (w_{a,i}^{(t+1)} - w_{a,i}^{(t)}) \phi_{T_{t+1}}(X_i).
\]
The two entries of each residual vector sum to zero, so its squared norm is twice its left-leaf discrepancy squared. Also $\|\phi_{T_{t+1}}(X_i)\|_2 = 1$. The triangle inequality and Cauchy--Schwarz therefore yield
\[
  \begin{aligned}
    S_{t+1} & \le 2G^2 \sum_a \|r_a(w_a^{(t)}; \phi_{T_{t+1}})\|_2^2\\
    & \le 4G^2 \sum_a \left\{\|r_a(w_a^{(t+1)}; \phi_{T_{t+1}})\|_2^2 + n_a\|w_a^{(t+1)} - w_a^{(t)}\|_2^2\right\}.
  \end{aligned}
\]
Thus a large selected score requires either residual imbalance after refitting or appreciable weight movement. The proof of Proposition~\ref{prop:bgf imbalance} bounds the comparator on the structure sample and averages the late iterations. The factor $n_a$ in the movement term is essential as this display uses probability weights, whereas that proof uses mean-one weights.

\lead{Construction for Figure~\ref{fig:weight feedback}.}
We place 100 observations at each point of $\{(i-0.5)/8: i = 1,\ldots, 8\}^2$, giving a structure sample of size $n=6400$. Write $h_c(z)=2 \cdot \one\{z<c\}-1$. At each point $x$, we assign $100 e(x)$ observations to arm $1$ and the remainder to arm $0$, where
\[
  e(x)=0.5+0.14h_{0.5}(x_1)+0.09h_{0.5}(x_2)+0.06h_{0.25}(x_1)h_{0.75}(x_2).
\]
These counts are integers, with $n_0=3296$ and $n_1=3104$. The empirical target law $Q_S$ is uniform on the 64 grid points. For both $a = 0,1$, let  $\widehat{\mu}_a$ be 
\[
  \begin{aligned}
    \widehat\mu_a(x)
    &=3 h_{0.5}(x_1) + 2 h_{0.5}(x_2) + 1.4 h_{0.25}(x_1) + 0.9 h_{0.75}(x_2)+0.8 h_{0.5}(x_1) h_{0.75}(x_2).
  \end{aligned}
\]

Trees are grown on the full structure sample with $m_{\mathrm{try}}=2$, $d_{\max}=2$, $m=400$, and $\alpha=0.05$. Both coordinates and all admissible thresholds $k/8$, $k=1,\ldots,7$, are considered at each node. The minimum child size is $\max\{400,\lceil 0.05 n(N)\rceil \}$, where $n(N)$ is the pooled observation count in the parent node. Starting from uniform weights within each arm, we grow each tree using \eqref{eq:ate score} and then refit both arms by \eqref{eq:ate refit}, with $\lambda_S=1$ and $C=4$. Each refit uses the leaf indicators from all $t$ trees, including repeated features, scaled by $t^{-1/2}$ after $t$ trees. $\widehat{\mu}_0$, $\widehat{\mu}_1$ and weights remain fixed during the growth of each tree.

For grid point $g$, let $u_{ag}^{(t)} = \sum_{i\in\mathcal{I}_a: X_i=g} w_{a,i}^{(t)}$ denote its weighted mass. Panel (a) displays $\rho_{1,t}(g) = 64 u_{1g}^{(t)} - 1$ for $t = 0,1,2,3$, with $\|\rho_{1,t}\|_n^2=64^{-1} \sum_g \rho_{1,t}(g)^2$. Panel (b) shows the first three trees and the dimensions of their cumulative leaf spaces, evaluated on the grid.

\subsection{How repetition changes the balancing metric}

For any tree $T$, exactly one leaf indicator is active at $x$, so $\|\phi_T(x)\|_2^2 = 1$. Hence $\|\Phi_t(x)\|_2^2 = k_t(x, x) = 1$ for every $t \ge 1$. Writing
\[
  D_T(w) = \sum_{\ell \in \mathcal{L}(T)} \{\widehat{P}_{a,w}(\ell) - Q_n(\ell)\}^2,
\]
the balance objective at step $t$ is $t^{-1} \sum_{b \le t} D_{T_b}(w)$. If $T$ already appears $k$ times, another copy changes its coefficient from $k/t$ to $(k + 1)/(t + 1)$, while each other occurrence has coefficient $1/(t + 1)$. Thus the linear span can stay fixed while the balancing metric changes. Even though feature spans accumulate, the associated coefficient-norm balls need not be nested. No monotonicity of MMD across forest sizes follows merely from adding trees.

\subsection{Changing the causal target}
\label{app:target extensions}

Fix the functions $b_a$ and the outcome transform $z$ in Section~\ref{sec:bgf extensions}. They may be learned on independent data, in which case the following statements are conditional on those data. Let $Q^\dagger \ll Q$, suppose $Q^\dagger \ll P_a$ for the arms being calibrated, and assume all displayed expectations are finite. Under Assumption~\ref{assump:identification},
\[
  \Psi = \sum_a \int b_a(x) \E\{z(Y(a)) \mid X = x\}\,dQ^\dagger(x) = \sum_a Q^\dagger g_a.
\]
For nonnegative weights with $P_a w_a = 1$, iterated expectation gives
\[
  \sum_a P_a \{w_a b_a z(Y)\} - \Psi = \sum_a (P_{a,w_a} - Q^\dagger)g_a.
\]
For a split with $Q^\dagger(L), Q^\dagger(R) > 0$, define $q^\dagger = Q^\dagger(L \mid N)$ and $g_{a,D} = Q^\dagger(g_a \mid D)$. The refinement is
\[
  h_{a,L;N}^\dagger = (g_{a,L} - g_{a,R})(\one_L - q^\dagger \one_N).
\]
Its target mean is zero, and its signed weighting discrepancy is
\[
  (P_{a,w_a} - Q^\dagger)h_{a,L;N}^\dagger = \{P_{a,w_a}(L) - q^\dagger P_{a,w_a}(N)\}(g_{a,L} - g_{a,R}).
\]
The sum of arm-specific squared contributions therefore gives the same BGF score with $Q$ replaced by $Q^\dagger$ and $\mu_a$ replaced by $g_a$. If $Q^\dagger f = Q(tf)/Qt$ for an available nonnegative tilt $t$ with $0 < Qt < \infty$, its empirical version is $Q_n(tf)/Q_n t$. Every target mean and node proportion uses this same tilt. A node with zero empirical target mass is not split. The ATT target instead uses the empirical covariate law of the treated arm directly.

For the ATT, $Q^\dagger = P_1$, $g_1 = \mu_1$, and $g_0 = -\mu_0$. On either the structure or analysis sample, uniform treated weights exactly match the empirical target feature mean and minimize the ridge penalty. They therefore solve the treated-arm calibration problem. With inherited uniform treated weights on a growth subsample, the within-parent treated discrepancy also vanishes whenever the empirical target child proportions are defined. All nonzero split contributions come from the control arm.

For a fixed policy $d(x) \in [0, 1]$, the mean under conditionally randomized treatment with probability $d(X)$ is $Q\{d \mu_1 + (1 - d) \mu_0\}$. Thus $b_1 = d$ and $b_0 = 1 - d$ give the estimator in \eqref{eq:extended estimator}. Within each node, average the functions $\widehat{g}_1=d \widehat{\mu}_1$ and $\widehat{g}_0 = (1 - d) \widehat{\mu}_0$ separately over the target covariates.

Finally, for a finite threshold grid and common final weights, define
\[
  \widehat{F}_a(u) = \sum_{i \in \mathcal{I}_a} \widehat{w}_{a,i} \one\{Y_i \le u\}.
\]
For $u < v$,
\[
  \widehat{F}_a(v) - \widehat{F}_a(u) = \sum_{i \in \mathcal{I}_a} \widehat{w}_{a,i} \one\{u < Y_i \le v\} \ge 0.
\]
Normalization gives limits zero and one, so each $\widehat{F}_a$ is a distribution function. Separate refits at each threshold would lose this automatic compatibility. The comparator argument for the target law $Q^\dagger$ requires $\|dQ^\dagger/dP_a\|_\infty \le C_\dagger < C$.

\section{Additional Experiments}
\label{app:additional experiments}

In this section, we describe the simulation design and additional experiments. Methods share the generated data within each replication.

\subsection{Simulation design and estimators}
\label{app:simulation details}

Each replication uses $3n$ independent observations, divided equally into nuisance, structure, and analysis samples $\mathcal{D}_N$, $\mathcal{D}_S$, and $\mathcal{D}_A$, without cross-fitting. Thus $n$ always denotes the analysis-sample size. Under the main design in Section~\ref{sec:experiments}, we use $n \in \{250, 500, 1000, 2000\}$, $p = 100$, and $\rho \in \{0, 0.5, 0.8\}$. At $n = 1000$, the construction paths use depths $d \in \{1, 2, 3, 5, 10, \infty\}$ and tree counts $B \in \{10, 20, 50, 100, 200, 400\}$. The depth comparison fixes $B = 400$. Sample-size comparisons use 400-tree BGF at depths one and two. Every configuration has 100 replications. The smooth, heterogeneous-effect, and pure-interaction experiments compare BGF(1), BGF(2), $K_Y$, $K_J$, OR, and RBF. The main sample-size comparison also includes nIPW and nAIPW. All experiments were run on a local machine with an AMD Ryzen 9 9955HX CPU and 64 GB of RAM.

\begin{table}[!htbp]
\centering
\caption{Training allocations and forest sizes. Estimation uses $\mathcal{D}_A$.}
\label{tab:simulation training}
\small
\begin{tabular}{@{}l>{\raggedright\arraybackslash}p{0.46\textwidth}>{\raggedright\arraybackslash}p{0.30\textwidth}@{}}
  \toprule
  Method & Training data and fitted components & Number of trees \\
  \midrule
  BGF, no-feedback
    & Outcome regressions on $\mathcal{D}_N$; shared forest on $\mathcal{D}_S$
    & 500 per nuisance arm; $B$ shared trees \\
  \addlinespace[2pt]
  $K_Y$
    & Separate outcome-guided forests in each arm of $\mathcal{D}_N \cup \mathcal{D}_S$
    & 1,000 total, 500 per arm \\
  \addlinespace[2pt]
  $K_J$ & Joint treatment--outcome forest on $\mathcal{D}_N \cup \mathcal{D}_S$ & 1,000 \\
  \addlinespace[2pt]
  OR & Arm-specific outcome regressions on $\mathcal{D}_N \cup \mathcal{D}_S$ & 500 per arm \\
  \addlinespace[2pt]
  RBF
    & Covariate scaling and bandwidth on $\mathcal{D}_N \cup \mathcal{D}_S$
    & Not applicable \\
  \addlinespace[2pt]
  nIPW & Propensity forest on $\mathcal{D}_N \cup \mathcal{D}_S$ & 1,000 \\
  \addlinespace[2pt]
  nAIPW
    & Outcome regressions on $\mathcal{D}_N$; propensity forest on $\mathcal{D}_S$
    & 500 per outcome arm; 1,000 propensity trees \\
  \bottomrule
\end{tabular}
\end{table}

All forests use 80\% subsampling without replacement and $m_{\mathrm{try}} = \max\{1, \lfloor p/3\rfloor\}$. BGF and the reference-kernel trees require at least $\max\{10, \lceil 0.05n(N) \rceil\}$ observations per child, where $n(N)$ is the number of growth-sample observations in the parent node $N$. BGF follows Algorithm~\ref{alg:bgf}, allowing admissible zero-score splits, holding weights fixed within each tree, and refitting them after each addition. The no-feedback ablation shares its nuisance fits and tree randomization but keeps the construction weights uniform.

The $K_Y$ kernel is adapted to ATE estimation by averaging the kernel given by the two arm outcome forests with variance splitting. The $K_J$ forest splits on the sum of variance reductions in $(A, Y)$ after standardizing both responses on the pooled training sample. Both reference kernels use unrestricted depth and the same final balancing problem as BGF. Outcome regression forests use variance splitting with a node-size stopping threshold of 10. OR averages $\widehat{\mu}_1(X_i) - \widehat{\mu}_0(X_i)$ over $\mathcal{D}_A$.

Forest kernels average the co-leaf indicators as in \eqref{eq:forest kernel}. For RBF, covariates are standardized using pooled training means and standard deviations, and 
\[
  k_h(z, z') = \exp\{ - \|z - z'\|^2/(2h^2)\}.
\]
The bandwidth maximizes the variance of off-diagonal kernel similarities on the training sample. Unless varied in Appendix~\ref{app:ridge sensitivity}, $\lambda_S = \lambda_A = 10^{-5}$ on the sum-to-one scale in \eqref{eq:ate refit}. All kernel-balancing weights satisfy $0 \leq w_{a,i} \leq 10/n_a$ and sum to one within arm, targeting the full empirical covariate distribution of the fitting sample. Final weights are fitted afresh on $\mathcal{D}_A$.

Propensity forests use Gini splitting and minimum node size 10, with training allocations in Table~\ref{tab:simulation training}. Let $\widehat{e}_{ai}$ be the fitted probability of treatment $a$ given $X_i$. For $i \in \mathcal{I}_{A,a} = \{i \in \mathcal{D}_A:A_i = a\}$, define
\[
  \widetilde{w}_{ai} = \frac{1/\widehat{e}_{ai}}{\sum_{j \in \mathcal{I}_{A,a}} 1/\widehat{e}_{aj}},
\]
with zero weights outside that arm. Then nIPW is $\sum_i (\widetilde{w}_{1i} - \widetilde{w}_{0i})Y_i$, and
\begin{equation}
  \widehat{\theta}_{\mathrm{nAIPW}} = \frac{1}{n} \sum_{i \in \mathcal{D}_A} \{\widehat{\mu}_1(X_i) - \widehat{\mu}_0(X_i)\} + \sum_{i \in \mathcal{D}_A} \widetilde{w}_{1i} \{Y_i - \widehat{\mu}_1(X_i)\} - \sum_{i \in \mathcal{D}_A} \widetilde{w}_{0i} \{Y_i - \widehat{\mu}_0(X_i)\}.
\end{equation}
Neither estimator clips propensities or caps weights.

All simulation designs have population ATE $\theta = 1$. For $R = 100$, let $E_r = \widehat{\theta}_r - 1$ and $\overline{E^2} = R^{-1} \sum_r E_r^2$. We report
\begin{equation}
  \widehat{\mathrm{RMSE}} = \sqrt{\overline{E^2}}, \qquad \widehat{\mathrm{MCSE}}(\widehat{\mathrm{RMSE}}) = \frac{\operatorname{sd}(E_1^2, \ldots, E_R^2)}{2 \widehat{\mathrm{RMSE}} \sqrt{R}}.
\end{equation}
Bands and error bars are $\widehat{\mathrm{RMSE}} \pm 1.96 \widehat{\mathrm{MCSE}}$, truncated at zero below. They describe Monte Carlo uncertainty in RMSE. Here $\operatorname{sd}$ is the sample standard deviation.

For each signal $k \in \{C, P, I\}$ in Section~\ref{sec:experiments}, let $\bar{g}_{k,r}$ and $s_{k,r}$ be its unweighted mean and sample standard deviation in the full analysis sample of replication $r$. With arm weights summing to one, define
\begin{equation}
  \delta_{a,k,r} = \frac{\sum_{i \in \mathcal{D}_A:A_i=a} \widehat{w}_{a,i} g_k(X_i) - \bar{g}_{k,r}}{s_{k,r}}, \qquad \mathrm{ASI}_{k,r} = \frac{|\delta_{0,k,r}| + |\delta_{1,k,r}|}{2}.
\end{equation}
Figure~\ref{fig:simulation}(d) averages ASI over the 100 replications at $n = 1000$. This measures imbalance in the aggregate nonlinear signals. The unweighted comparison uses uniform within-arm weights.

\subsection{Minimum child size in the reference forests}
\label{app:minimum child size}

\begin{figure}[!htbp]
\centering
\includegraphics[width=\textwidth]{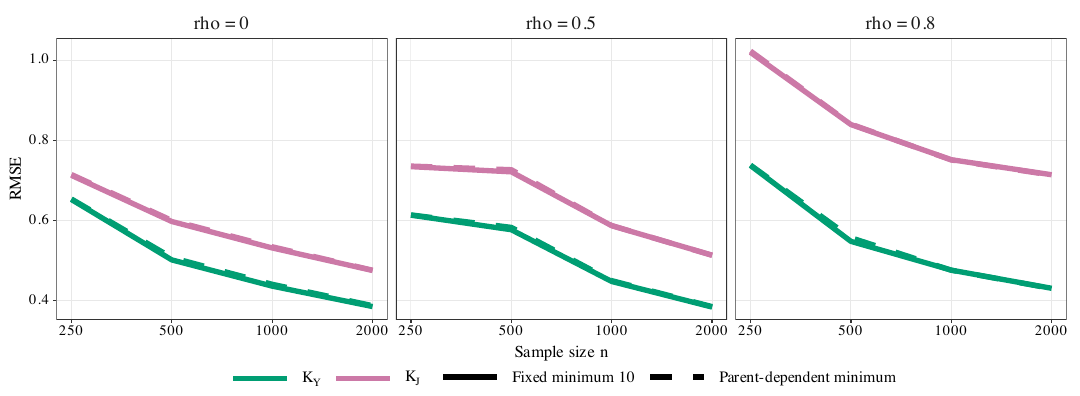}
\caption{Minimum-child sensitivity for $K_Y$ and $K_J$.}
\label{fig:minimum child size}
\end{figure}

We compare the fixed minimum of 10 observations per child with the default $\max\{10, \lceil0.05n(N)\rceil\}$ for $K_Y$ and $K_J$ across the main design's sample sizes and correlations. Each uses 1,000 trees and 100 replications, with matched data and forest randomization. Figure~\ref{fig:minimum child size} shows nearly overlapping RMSE curves, indicating little sensitivity to this choice.

\subsection{Smooth nonlinear features}
\label{app:smooth features}

This experiment replaces the absolute-value signals of the main design with smooth nonlinear functions and an interaction. We generate independent $X_j \sim N(0, 1)$ for $j= 1, \ldots, 100$, and define
\[
  C(X) = \sum_{j=1}^{6} \{\cos(X_j) - e^{-1/2}\}, \qquad H(X) = \sin(X_1 + X_2), \qquad P(X) = \sum_{j=7}^{12} \{\cos(X_j) - e^{-1/2}\}.
\]
Treatment and outcome follow
\[
  \begin{aligned}
    e(X) & = \operatorname{expit}\{0.25C(X) + 0.5H(X)\}, \\
    A \mid X & \sim \operatorname{Bernoulli}\{e(X)\}, \\
    Y & = C(X) + 2H(X) + 2P(X) + A + \varepsilon,
  \end{aligned}
\]
where $\varepsilon \sim N(0, 1)$ is independent of $(X, A)$. The true ATE is one.

\begin{figure}[!htbp]
\centering
\includegraphics[width=0.6\textwidth]{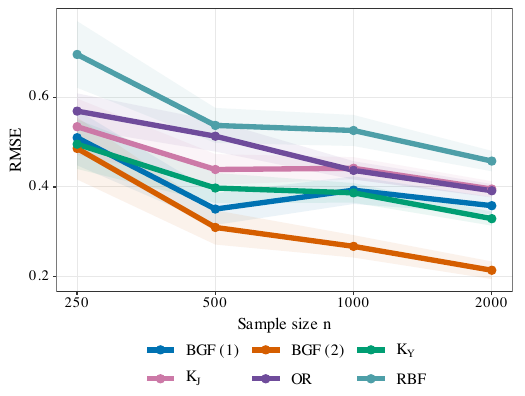}
\caption{ATE RMSE in the smooth-feature design with $p = 100$ independent Gaussian covariates. Bands show approximate 95\% Monte Carlo intervals.}
\label{fig:smooth features}
\end{figure}

\subsection{Heterogeneous treatment effects}
\label{app:heterogeneous effects}

To assess ATE estimation when the treatment effect varies across observations, we generate independent $X_j \sim N(0, 1)$, $j = 1, \ldots, 100$, and let
\[
  q_j = \mathbf{1}\{|X_j| > 1\} - 2 \Phi(-1), \qquad H(X) = q_1 q_2 + q_3 q_4,
\]
where $\Phi$ is the standard normal distribution function. Treatment is Bernoulli with probability
\[
  e(X) = 0.1 + 0.8 \operatorname{expit}\!\left[0.25 \sum_{j=1}^{6} \left\{|X_j| - \sqrt{2/\pi}\right\} + 0.5H(X)\right],
\]
and the outcome is
\[
  Y = 2 \sum_{j=1}^{6} |X_j| + 2H(X) + A(1 + 2q_5 q_6) + \varepsilon, \qquad \varepsilon \sim N(0, 1), \qquad \varepsilon \perp (X, A).
\]
The conditional effect is $\tau(X) = 1 + 2 q_5 q_6$. Since $q_5$ and $q_6$ are independent and mean zero, the true ATE remains $\mathbb{E}\{\tau(X)\} = 1$. We use the same four sample sizes and six methods as in the smooth experiment.

\begin{figure}[!htbp]
\centering
\includegraphics[width=0.6\textwidth]{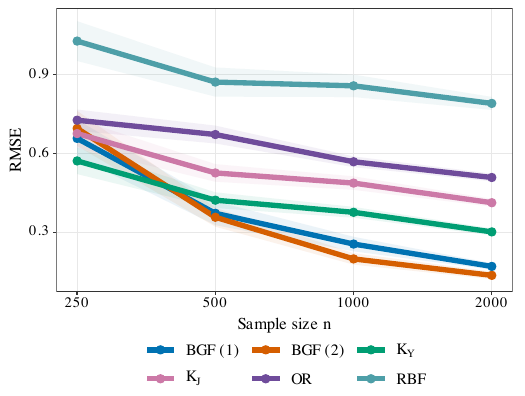}
\caption{ATE RMSE under heterogeneous treatment effects $\tau(X) = 1 + 2q_5 q_6$. Bands show approximate 95\% Monte Carlo intervals.}
\label{fig:heterogeneous effects}
\end{figure}

\subsection{Construction-ridge sensitivity}
\label{app:ridge sensitivity}

\begin{figure}[!htbp]
\centering
\includegraphics[width=\textwidth]{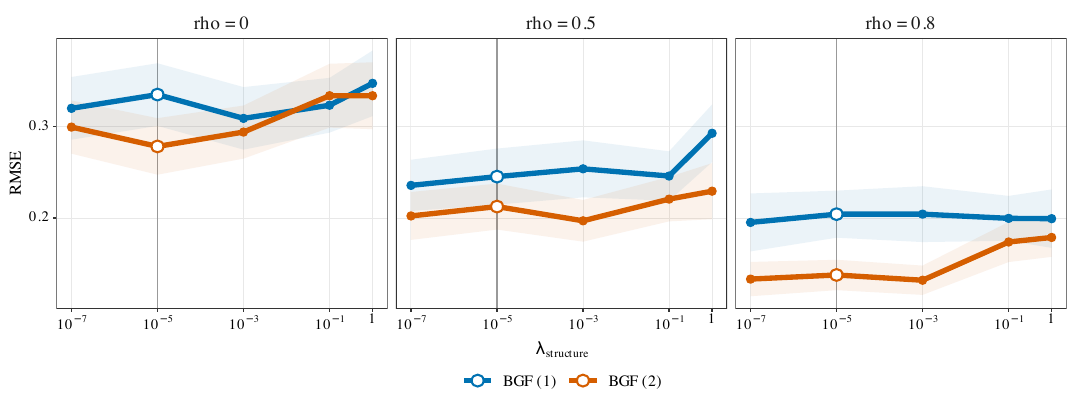}
\caption{Sensitivity to the construction ridge $\lambda_S$, with the analysis ridge fixed at $\lambda_A = 10^{-5}$. The hollow circles mark the default $\lambda_S = 10^{-5}$. Bands show approximate 95\% Monte Carlo intervals for RMSE.}
\label{fig:ridge sensitivity}
\end{figure}

In this section, we explore the ridge sensitivity of shallow BGF methods on the structure sample. We vary $\lambda_S \in \{10^{-7}, 10^{-5}, 10^{-3}, 10^{-1}, 1\}$ while fixing $\lambda_A = 10^{-5}$. The main design is used with $n = 1000$, $p = 100$, all three correlation settings, and 400 trees. Each configuration has 100 replications, with data, nuisance fits, and forest randomization matched across ridge values. The forest is rebuilt at each $\lambda_S$. Figure~\ref{fig:ridge sensitivity} supports local stability near the default.

\subsection{Pure-interaction stress test}
\label{app:pure interaction stress test}

This experiment tests the dependence of shallow BGF on marginal outcome guidance. We generate $X \sim N_{15}(0, \Sigma_\rho)$, with $(\Sigma_\rho)_{jk} = \rho^{|j-k|}$ and $\rho \in \{0, 0.5, 0.8\}$. Let $U = \operatorname{sign}(X_1 X_2)$ and generate
\[
  e(X) = \begin{cases}
    3/4, & X_1 X_2 > 0, \\
    1/4, & X_1 X_2 \leq 0,
  \end{cases}
  \qquad A \mid X \sim \operatorname{Bernoulli}\{e(X)\}, \qquad Y = U + A + \varepsilon,
\]
where $\varepsilon \sim N(0, 1)$ is independent of $(X, A)$. Thus $Y(0) = U + \varepsilon$, $Y(1)= 1 + U + \varepsilon$, and the ATE is one. We use the same four sample sizes, six methods, and 100 replications as above. With $p = 15$, the candidate-variable count becomes $m_{\mathrm{try}}= 5$.

When $\rho = 0$ and $\widehat{\mu}_a = \mu_a$ for $a= 0, 1$, $\mathbb{E}(U \mid X_j) = 0$ for every coordinate. Every root-level population outcome contrast is therefore zero. Stumps receive no marginal guidance toward the sign interaction, although depth-two trees can represent it.

\begin{figure}[htbp]
\centering
\includegraphics[width=0.98\textwidth]{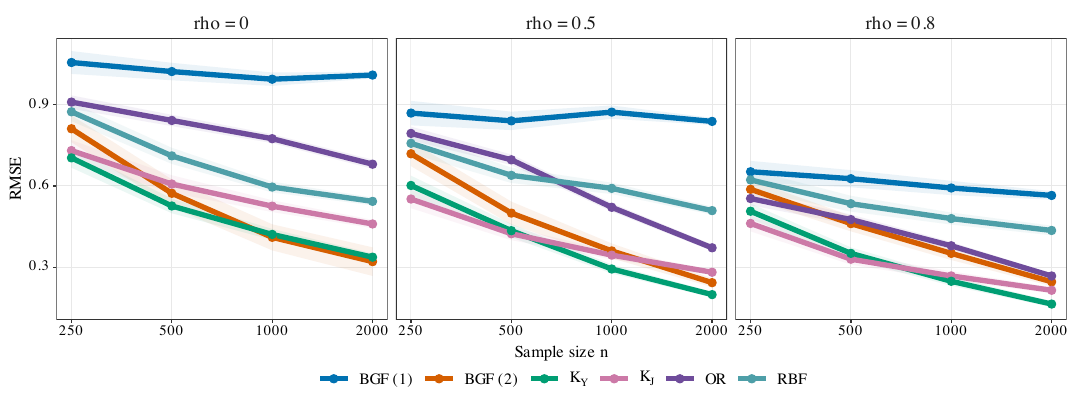}
\caption{Pure-interaction stress test. Bands show approximate 95\% Monte Carlo intervals.}
\label{fig:pure interaction}
\end{figure}

Figure~\ref{fig:pure interaction} reflects this limitation. At $\rho = 0$, BGF(1) has RMSE close to one throughout, while the performance of BGF(2) is improving as sample size increases. Access to interactions and the quality of marginal guidance constrain the benefit of shallow feedback.

\FloatBarrier
\section{Real-data Applications}
\label{app:real data implementation}

\subsection{Data and adjustment variables}
\label{app:real data preprocessing}

Table~\ref{tab:real data samples} summarizes the samples used by all methods. Categorical covariates use treatment contrasts without an intercept. RBF scaling uses only training folds.

\begin{table}[htbp]
\centering
\caption{Real-data samples with total size $N$, treated count $N_1$, encoded covariate dimension $p$, and outcome scale.}
\label{tab:real data samples}
\begin{tabular}{lrrrl}
  \toprule
  Dataset & $N$ & $N_1$ & $p$ & Outcome and reporting scale \\
  \midrule
  RHC   & 5,735 & 2,184 & 65 & 30-day mortality; percentage-point risk difference \\
  NHEFS & 1,566 & 403   & 14 & Weight change; kilograms \\
  STAR  & 1,911 & 905   & 26 & Mathematics, reading, language; test-score points \\
  \bottomrule
\end{tabular}
\end{table}

\paragraph{RHC.}
We use Vanderbilt University's public RHC data,\footnote{Data obtained from \url{http://hbiostat.org/data} courtesy of the Vanderbilt University Department of Biostatistics.} with receipt of RHC within 24 hours of admission as treatment and death within 30 days as the binary outcome. Following \citet{de2025}, we estimate the target-population risk difference for 30-day mortality. We use 50 adjustment variables covering demographics, diagnoses and medical history, illness severity, and physiological and laboratory measurements, yielding 65 encoded covariates.

\paragraph{NHEFS.}
We use the complete-case version of the NHEFS data described by \citet{hernan2020}, available as \texttt{nhefs\_complete} in the R package \texttt{causaldata}. The treatment is smoking cessation, and the outcome is weight change from 1971 to 1982, measured in kilograms. Following \citet{ertefaie2023}, we adjust for sex, race, age, education, smoking intensity, years of smoking, exercise, activity level, and baseline weight. Education, exercise, and activity level are encoded using reference-category indicators, yielding 14 encoded covariates.

\paragraph{STAR.}
We use the Project STAR analysis sample described by \citet{imai2023}, available as \texttt{star} in the R package \texttt{evalITR}. Treatment is small-class assignment relative to regular-class assignment. Outcomes are third-grade mathematics, reading, and language scores. Adjustment uses gender, race, birth month and year, and six school characteristics (\texttt{SCHLURBN}, \texttt{GKENRMNT}, \texttt{GRDRANGE}, \texttt{GKFRLNCH}, \texttt{GKBUSED}, and \texttt{GKWHITE}), using the categorical definitions in the source data. All three outcomes use the same students. Results appear in Appendix~\ref{app:star}.

\subsection{Cross-fitting and estimation}
\label{app:real data crossfit}

We repeat treatment-stratified three-fold cross-fitting five times, rotating folds cyclically through $\mathcal{D}_N$, $\mathcal{D}_S$, and $\mathcal{D}_A$. Methods share partitions and follow Table~\ref{tab:simulation training}. STAR outcomes share partitions but have separately fitted outcome-dependent forests.

All applications use the simulation settings in Appendix~\ref{app:simulation details}: 400-tree BGF at depth one or two, $\lambda_S = \lambda_A = 10^{-5}$, and kernel-weight cap $C = 10$. For analysis-fold estimate $\widehat{\theta}_{rf}$ and fold size $N_{rf}$, the reported adjusted estimate is
\begin{equation}
  \widehat{\theta}_{\mathrm{CF}} = \frac{1}{5} \sum_{r=1}^{5} \sum_{f=1}^{3} \frac{N_{rf}}{N} \widehat{\theta}_{rf}.
\end{equation}
EmpDiff is the full-sample simple difference in means between the treatment group and the control group. We report point estimates.

\subsection{Covariate-balance diagnostics}
\label{app:real data balance}

Let $s_j = \{(s_{0j}^2 + s_{1j}^2)/2\}^{1/2}$, where $s_{aj}^2$ is encoded covariate $j$'s unweighted within-arm sample variance in the full dataset. With this fixed denominator and weights normalized within analysis-fold arms, define
\begin{equation}
  \mathrm{SMD}_{j,rf} = \frac{\bar{X}^w_{1j,rf} - \bar{X}^w_{0j,rf}}{s_j}, \qquad D_j = \frac{1}{15} \sum_{r=1}^{5} \sum_{f=1}^{3} |\mathrm{SMD}_{j,rf}|.
\end{equation}
Here $\bar{X}^w_{aj,rf}$ is the weighted mean of covariate $j$ in arm $a$ of the analysis fold indexed by repetition $r$ and fold $f$. The plotted ECDF is $p^{-1} \sum_j \mathbf{1}\{D_j \leq t\}$. Upper-left curves indicate smaller marginal imbalance. Absolute SMDs are averaged equally over the 15 folds. The unadjusted curve uses uniform within-arm weights. OR has no balancing weights, and nAIPW is also omitted as it is not considered as a balancing method. 

\subsection{Project STAR}
\label{app:star}

Project STAR randomized kindergarten class size \citep{mosteller1995}. Following \citet{imai2023}, we analyze the small-versus-regular-class comparison described in Appendix~\ref{app:real data preprocessing}. For all methods except EmpDiff, we repeat cross-fitting as described above.

Table~\ref{tab:star ate} gives positive point estimates for all methods and outcomes.

\begin{table}[!htbp]
\centering
\caption{STAR ATE estimates for small-class assignment, in score points.}
\label{tab:star ate}
\setlength{\tabcolsep}{16pt}
\begin{tabular}{lrrr}
  \toprule
  Method & Mathematics & Reading & Language \\
  \midrule
  EmpDiff & 5.777 & 6.777 & 3.648 \\
  BGF(1)  & 5.436 & 5.776 & 2.502 \\
  BGF(2)  & 4.313 & 6.044 & 2.664 \\
  $K_Y$   & 4.724 & 5.601 & 2.793 \\
  $K_J$   & 4.834 & 5.759 & 2.829 \\
  OR      & 5.877 & 6.055 & 3.192 \\
  RBF     & 6.932 & 7.558 & 4.399 \\
  nIPW    & 5.891 & 6.319 & 3.424 \\
  nAIPW   & 6.712 & 7.214 & 4.232 \\
  \bottomrule
\end{tabular}
\end{table}

\begin{figure}[!htbp]
\centering
\includegraphics[width=0.98\textwidth]{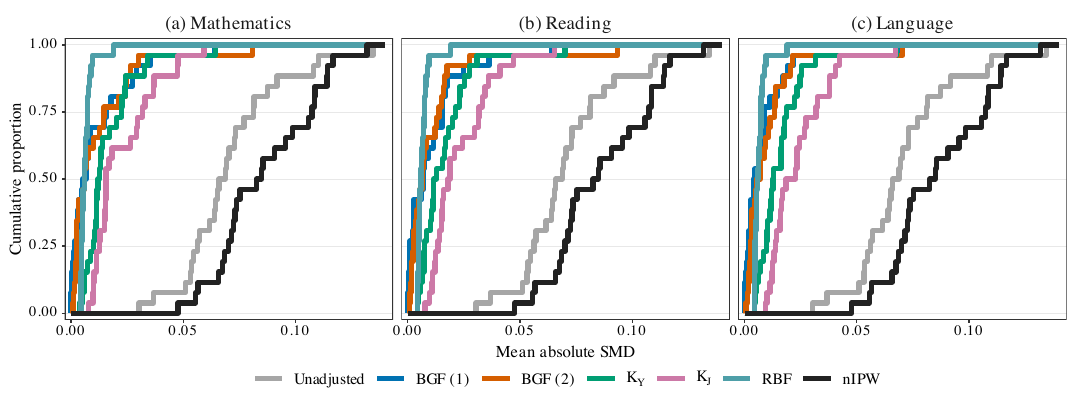}
\caption{Covariate-balance ECDFs for STAR. Upper-left curves indicate better balance.}
\label{fig:star balance}
\end{figure}

Figure~\ref{fig:star balance} shows reduced marginal imbalance for RBF and BGF relative to the unadjusted comparison.

\end{document}

%% file: layout.tex
\usepackage[
  letterpaper,
  left=1.20in,right=1.20in,
  top=1in,bottom=1in,
  headheight=14pt,headsep=16pt,footskip=25pt
]{geometry}
\usepackage{microtype}

\usepackage{indentfirst}
\usepackage{titlesec}
\usepackage{needspace}
\usepackage{etoolbox}
\usepackage{caption}
\usepackage{fancyhdr}
\usepackage{xurl}

\renewcommand{\footnotesize}{\small}
\newcommand{\BGFfootnotefont}{\linespread{1}\fontsize{12}{15}\selectfont}
\makeatletter
\patchcmd{\@footnotetext}{\footnotesize}{\BGFfootnotefont}{}{%
  \PackageError{compact-layout}{Footnote font patch failed}{%
    The standard article footnote command was not found.}%
}
\makeatother

\titleformat{\section}[runin]
  {\normalfont\normalsize\bfseries}{\thesection.}{0.6em}{}[.]
\titlespacing*{\section}{\parindent}{11pt plus 2pt minus 2pt}{0.7em}
\titleformat{name=\section,numberless}[block]
  {\normalfont\normalsize\scshape\centering}{}{0pt}{}
\titlespacing*{name=\section,numberless}{0pt}{11pt plus 2pt minus 2pt}{7pt}
\titleformat{\subsection}[runin]
  {\normalfont\normalsize\itshape}{\thesubsection.}{0.6em}{}[.]
\titlespacing*{\subsection}{\parindent}{8pt plus 2pt minus 1pt}{0.6em}
\titleformat{\subsubsection}[runin]
  {\normalfont\normalsize\itshape}{\thesubsubsection.}{0.6em}{}[.]
\titlespacing*{\subsubsection}{\parindent}{7pt plus 1pt minus 1pt}{0.6em}
\titleformat{\paragraph}[runin]
  {\normalfont\normalsize\bfseries}{}{0pt}{}
\titlespacing*{\paragraph}{\parindent}{5pt plus 1pt minus 1pt}{0.5em}

\pretocmd{\section}{\Needspace{4\baselineskip}}{}{}
\pretocmd{\subsection}{\Needspace{3\baselineskip}}{}{}

\newtheoremstyle{compactplain}
  {5pt plus 1pt minus 1pt}{5pt plus 1pt minus 1pt}
  {\itshape}{\parindent}{\scshape}{.}{0.5em}
  {\thmname{#1}\thmnumber{ #2}\thmnote{ \normalfont(#3)}}
\newtheoremstyle{compactremark}
  {5pt plus 1pt minus 1pt}{5pt plus 1pt minus 1pt}
  {\normalfont}{\parindent}{\scshape}{.}{0.5em}
  {\thmname{#1}\thmnumber{ #2}\thmnote{ \normalfont(#3)}}

\allowdisplaybreaks[1]
\AtBeginDocument{%
  \setlength{\abovedisplayskip}{6pt plus 2pt minus 1pt}%
  \setlength{\belowdisplayskip}{6pt plus 2pt minus 1pt}%
  \setlength{\abovedisplayshortskip}{3pt plus 1pt}%
  \setlength{\belowdisplayshortskip}{4pt plus 1pt minus 1pt}%
}
\setlist{topsep=4pt,itemsep=2pt,parsep=0pt}

\fancypagestyle{plain}{%
  \fancyhf{}%
  \fancyfoot[C]{\footnotesize\thepage}%
}

\makeatletter
\renewcommand{\@maketitle}{%
  \vspace*{20pt}%
  \begin{center}%
    {\fontsize{14}{18}\selectfont\bfseries\MakeUppercase{\@title}\par}%
    \vskip 11pt
    {\normalsize\@author\par}%
  \end{center}%
  \vskip 3pt
}
\renewenvironment{abstract}{%
  \small
  \list{}{%
    \setlength{\leftmargin}{0.60in}%
    \setlength{\rightmargin}{0.60in}%
    \setlength{\listparindent}{1.25em}%
    \setlength{\itemindent}{1.25em}%
    \setlength{\topsep}{4pt}%
    \setlength{\parsep}{0pt}%
  }%
  \item\relax\textsc{\abstractname.}\enspace\ignorespaces
}{%
  \endlist
}
\makeatother